\documentclass{article}
\usepackage{graphicx} 
\usepackage{amsmath}
\usepackage{geometry}
\usepackage{bm}
\usepackage{subfigure}
\usepackage{color}
\usepackage{authblk}

\usepackage[title]{appendix}

\title{A novel boundary-integral algorithm for nonlinear unsteady surface and interfacial waves}
\author[1]{Xin Guan\thanks{xin.guan.20@ucl.ac.uk}}
\author[1]{Jean-Marc Vanden-Broeck}
\affil[1]{Department of Mathematics, University College London, London WC1E 6BT, UK}
\date{}

\begin{document}

\maketitle

\begin{abstract}
    \noindent We devise a new time-stepping algorithm for two-dimensional nonlinear unsteady surface and interfacial waves. The algorithm uses Cauchy's integral formula, which only requires information on the interface, to solve Laplace equation by using iterative techniques. We derive Eulerian and mixed Eulerian-Lagrangian descriptions by using arclength to parameterize the interface which is updated through its inclination angle and velocity potential at each time step. The algorithm shows broad applicability and excellent numerical accuracy in various numerical simulations, including wave breaking, collisions of solitary waves, vortex roll-up, etc. We especially focus on the stability of symmetric interfacial gravity-capillary solitary waves in deep water. Linear stability analysis is performed using a new formulation which possesses excellent numerical efficiency and robustness. It is shown that the depression/elevation solitary waves are linearly stable/unstable except the portion where the monotonicity of energy curve changes firstly. These results are supported by our fully nonlinear simulations, especially the head-on collisions of solitary waves.
\end{abstract}

\section{Introduction}
Waves on the interfaces of two immersible fluids are known as interfacial waves. As an extension of water waves, they can date back to Kelvin and Helmholtz in their studies of hydrodynamic instabilities. Many physical related scenarios can be effectively modelled by interfacial waves, for instance, air-water interactions, oceanic internal waves, etc. These phenomena are usually highly nonlinear and can not be well described by weakly nonlinear models, such as the Korteweg–de Vries equation or the nonlinear Schr\"{o}dinger (NLS) equation, leaving nonlinear simulations to be the only approach in many cases.

As a special case of interfacial waves, water waves have been numerically simulated for decades and proven to be an active and fruitful area. Under the three classical assumptions, i.e. the flow is incompressible, irrotational, and inviscid, the Euler equations are equivalent to the Laplace equation with boundary conditions on a time-dependent surface. This enables to use numerical methods of boundary-integral type, which have advantages of reducing the dimensionality of the original problems. In two dimensions, these numerical algorithms can be approximately separated into three kinds: (1) Green's function method based on Green's identities \cite{longuet1976deformation,yang1992initial}, (2) Cauchy's integral method based on Cauchy's integral formula for analytic functions \cite{dold1992efficient,murashige2017numerical}, and (3) vortex method based on singularity distributions and the Biot-Savart integral \cite{baker1983generalized}. Sample points are initially spread on the surface and labelled by a parameter $\xi$. Their physical coordinates $\vec r(\xi,t) = \big(x(\xi,t),\eta(\xi,t)\big)$ are tracked during the simulation. This is the essence of the Lagrangian description, which is popular due to its simple mathematical formulation. The particle-tracking strategy offers adaptability, allowing sample particles to cluster in regions of high curvature, which is especially attractive when studying breaking waves \cite{longuet1976deformation}. On the other hand, the Eulerian description fixes the positions of sample points in the physical space, e.g. their $x$-coordinates, or in some transformation spaces \cite{murashige2017numerical,ribeiro2017flow}, usually leading to complicated formulations.

When it comes to two-fluid system, adopting the same assumptions as in the water waves gives benefits to simplifying the mathematical formulation of interfacial waves, especially by allowing the application of boundary-integral methods. The inviscid assumption leads to a discontinuity in tangential velocity across the interface. Since flows are irrotational in the interior of fluid, vorticity is confined to the interface, making it a vortex sheet with zero thickness. Therefore, the vortex method is especially popular in studies of interfacial motions \cite{baker1980vortex,pullin1982numerical,hou1994removing,hou2001boundary}, including wave propagation, instabilities of the Rayleigh-Taylor (R-T) and Kelvin-Helmholtz (K-H) type, etc. Boundary-integral methods based on Cauchy's integral formula, although less commonly used, have advantages in situations where topography plays a crucial role. To our knowledge, the only application in unsteady interfacial waves via this approach is \cite{grue1997method} where the authors investigated unsteady transcritical two-layer flow over a bottom topography. Their method is a direct extension of that in \cite{dold1992efficient} for surface waves, based on the Lagrangian description.

In many situations, surface tension is important for interfacial waves, either physically relevant or as a regularization of the K-H singularity. In this context, Eulerian description or mixed Eulerian-Lagrangian description are better choices because Lagrangian description may lead to an overly sparse distribution of sample particles, resulting in poor physical resolution. In \cite{baker1980vortex}, the authors found that sample particles tend to move away from the developing spike and bubble regions, thus restricting the numerical accuracy. In addition, the accumulation of particles increases local wave number and may cause severe numerical stiffness, as reported in \cite{hou1994removing}. A natural parameterization for Eulerian description is to use the arclength $s$, which guarantees a uniform distribution of sample points on the interface regardless of the topology of waves. The physical coordinates of these points, as well as their velocity potential values are the unknowns to be solved,  leading to a discretized system with $O(3N)$ unknowns, where $N$ is the number of sample points on the interface. Alternatively, a more convenient approach is to use the inclination angle $\theta$ of the interface as unknown. This only requires $O(2N)$ unknowns and leads to a $\theta-s$ formulation. 

In this paper, we describe a new boundary-integral algorithm to simulate two-dimensional nonlinear interfacial waves and surface waves. The basic idea is to use a pair of functions: $\theta$ and $\varphi$, to describe the motion of interface or surface, where $\varphi$ is a density-weighted velocity potential defined in section \ref{section for phi}. To integrate these unknowns in time, normal velocity $\mathcal N$ on the interface is required. Given $\theta$ and $\varphi$, $\mathcal N$ is determined from Cauchy's integral formula, which is reformulated into a Fredholm integral equation of the second kind. This equation is solved iteratively by using the generalized minimal residual method (GMRES). Note that this idea is analogue to the Hamiltonian formulation of water waves and interfacial waves that uses $\eta$ (surface elevation) and $\varphi$ as a pair of canonical variables and Dirichlet-to-Neumann operator to solve $\mathcal N$ \cite{zakharov1968stability,benjamin1997reappraisal}. To derive an Eulerian description, we adopt an arclength-parameterization method to ensure uniform spacing of sample points along the interface or surface. It is worth noting that two other works have employed a very similar concept. In \cite{yang1992initial}, the author used a combination of $\theta-s$ formulation and Green's function to study  
a fluid falling into vacuum. In \cite{hou1994removing}, $\theta-s$ formulation coupled with vortex sheet method is employed to study Hele-Shaw flow and vortex roll-up structures due to the K-H instability.

Using the combination of $\theta-s$ formulation and Cauchy's integral formula, we are able to simulate various wave phenomena, such as wave breaking, collision of solitary waves, vortex roll-up, etc. Particularly, we focus on the stability of symmetric interfacial gravity-capillary solitary waves in deep water. To our knowledge, there are few works on this topic due to the lack of good algorithms for unsteady interfacial waves. In \cite{calvo2002stability}, the authors studied the linear stability of surface gravity-capillary solitary waves in deep water. There are two kinds of symmetric solitary waves which feature negative and positive values of $\eta$ on their center in small amplitude, known as depression and elevation solitary waves (see Fig. \ref{fig:bif1} and \ref{fig:ele_bif}). It was found that the depression solutions are linearly stable. On the other hand, the elevation solutions are linearly unstable until a stationary point of energy emerges on their bifurcation. Subsequently, solutions become linearly stable until a second stationary appears. These findings are supported by the theory of exchange of stability \cite{saffman1985superharmonic} in superharmonic case and various nonlinear numerical experiments \cite{milewski2010dynamics,wang2016stability}. In \cite{calvo2003interfacial}, the authors extended the linear stability analysis to interfacial gravity-capillary solitary waves in a shallow water-deep water setting and found that the linear stability is density-dependent. To our knowledge, this is the only work regarding the linear stability of interfacial gravity-capillary solitary waves. We have used the same method derived in \cite{calvo2003interfacial} but encountered some numerical issues. The problem is intrinsic to their formulation which has a form of generalised eigenvalue problems: $Ax = \lambda Bx$, where $A$ and $B$ are $3N\times3N$ matrices, and especially $B$ is singular. To address this, we reformulate the analysis, partly inspired by the $\theta-s$ formulation, by introducing new variables. This ultimately gives rise to a $2N\times 2N$ eigenvalue problem in standard form, possessing both numerical robustness and efficiency. This formulation are employed to interfacial gravity-capillary solitary waves in deep water, for a fixed density ratio $0.2$. It is found that for depression solitary waves, their energy bifurcation has two stationary points and the solutions between them are linearly unstable. For elevation solitary waves, their linear stability properties are almost same to that were previously found for surface elevation solitary waves, except there is a dominant superharmonic instability before the second exchange of stability happens. These results are supported by our fully nonlinear simulations that use perturbed solitary waves as the initial conditions and monitor the growth of the disturbances. We also perform head-on collisions between solitary waves, which yield the same conclusions on their stability. It is worth mentioning that we found our time-stepping algorithm is numerically robust and very accurate in long-term simulations. 

\section{Mathematical formulation}
\begin{figure}[h!]
    \centering
    \includegraphics[width = 0.65\textwidth]{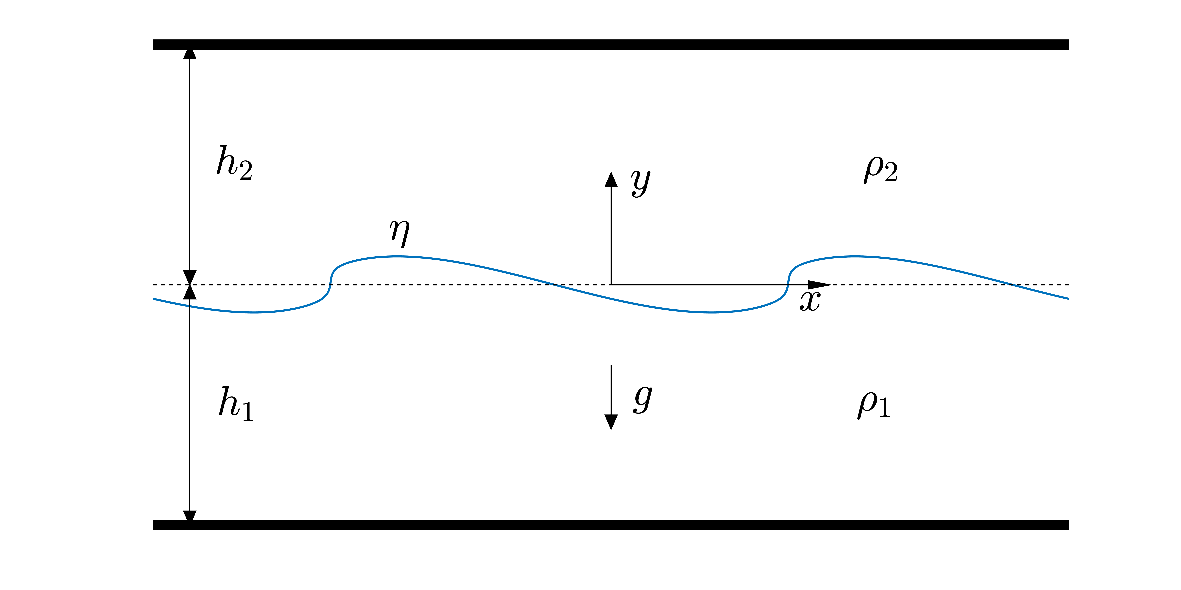}
    \caption{A schematic of the flow configuration.}
    \label{fig:schematic}
\end{figure}

We consider a system composed of two immersible, incompressible and inviscid fluids with different densities (see Fig. \ref{fig:schematic}). The lighter fluid lies above the heavier one to keep a linearly stable configuration. Inside each fluid, we assume that the motion is two-dimensional and irrotational. Thus we can introduce two velocity potential functions $\phi_{i}$, where subscripts $i=1,2$ are used to represent properties of the the lower and upper fluid. We also assume that each fluid has uniform depth $h_i$ and constant density $\rho_i$. Thus we can write down the Laplace equation in each fluid layer
\begin{align}
    \phi_{1,xx}+\phi_{1,yy} &= 0, \qquad -h_1<y<\eta,\label{Laplace1}\\
    \phi_{2,xx}+\phi_{2,yy} &= 0, \qquad \eta<y<h_2,\label{Laplace2}
\end{align}
where we have used $\eta(x,t)$ to denote the elevation of the interface and put the $x$-axis on the mean water level.
On the interface, two kinematic boundary conditions and a dynamic boundary condition are imposed
\begin{align}
    \eta_t +& \phi_{1,x}\eta_x - \phi_{1,y} = 0,\label{kinematic1}\\
    \eta_t +& \phi_{2,x}\eta_x - \phi_{2,y} = 0,\label{kinematic2}\\
    \rho_1 \phi_{1,t} - \rho_2 \phi_{2,t} + \frac{\rho_1}{2}\big(\phi_{1,x}^2+\phi_{1,y}^2\big) -& \frac{\rho_2}{2}\big(\phi_{2,x}^2+\phi_{2,y}^2\big)  + (\rho_1-\rho_2)g\eta - \sigma \frac{\eta_{xx}}{(1+\eta_x^2)^{3/2}} = 0,\label{dynamic1}
\end{align}
where $g$ is the acceleration due to the gravity, and $\sigma$ is the coefficient of surface tension. On the top and bottom wall, we impose two impermeability boundary conditions
\begin{align}
    \phi_{1,y} &= 0, \quad y = -h_1,\label{imper1}\\
    \phi_{2,y} &= 0, \quad y = h_2.\label{imper2}
\end{align}
By choosing 
\begin{align}
    \bigg(\frac{\sigma}{\rho_1 g}\bigg)^{1/2},\qquad \bigg(\frac{\sigma g}{\rho_1}\bigg)^{1/4},\qquad \bigg(\frac{\sigma}{\rho_1 g^3}\bigg)^{1/4}
\end{align}
as typical length, speed and time scales, we make the above formulation dimensionless. Eqs. (\ref{Laplace1})-(\ref{kinematic2}), (\ref{imper1}), and (\ref{imper2}) are invariant. The dynamic boundary condition (\ref{dynamic1}) now reads
\begin{align}
    \phi_{1,t} - R \phi_{2,t} + \frac{1}{2}\big(\phi_{1,x}^2+\phi_{1,y}^2\big) - \frac{R}{2}\big(\phi_{2,x}^2+\phi_{2,y}^2\big)  + (1-R)\eta - \frac{\eta_{xx}}{(1+\eta_x^2)^{3/2}} = 0,\label{dynamic2}
\end{align}
where $R=\rho_2/\rho_1<1$ represents the density ratio. 

Linearizing the system and assuming solutions are of the following forms,
\begin{align}
    \eta(x,t) &= a_1\mathrm e^{\mathrm i (kx-\omega t)} + c.c.,\\
    \phi_1(x,y,t) &= \big(a_2\mathrm e^{ky} + a_3\mathrm e^{-ky}\big)\mathrm e^{\mathrm i (kx-\omega t)} + c.c.,\\
    \phi_2(x,y,t) &= \big(a_4\mathrm e^{ky} + a_5\mathrm e^{-ky}\big)\mathrm e^{\mathrm i (kx-\omega t)} + c.c.,
\end{align}
where $c.c.$ denotes complex conjugate, and $a_1$ to $a_5$ are unknown constants. Solving these coefficients, we can get the dispersion relation 
\begin{align}
    \omega^2 = \frac{(1-R)k+k^3}{\coth(k h_1)+R\coth(k h_2)},\label{c_p1}
\end{align}
where $\omega$ and $k$ represent angular frequency and wave number of linear waves respectively. Since $R<1$, the right hand side of (\ref{c_p1}) is always non-negative, thus we have a linearly stable system. If we let $h_{1,2}\rightarrow \infty$, we get the dispersion relation for deep-water waves
\begin{align}
    \omega^2 = \frac{(1-R)|k|+|k|^3}{1+R}\label{c_p}.
\end{align}

\section{$\theta-s$ formulation}       
\subsection{Interfacial waves}
Our aim is to derive a formulation for unsteady interfacial waves using the Eulerian or mixed Eulerian-Lagrangian description. It is natural to use arclength $s$, instead of the $x$-coordinate, to parameterize the interface since it could develop overhanging profiles. Without loss of generality, we assume the range of $s$ is $[0, S(t)]$, where $S(t)$ is the total arclength. For the convenience of discretization, we introduce the pseudo-arclength $l = s/S(t)$ whose range is always $[0,1]$. The motion of a Lagrangian particle on the interface is determined by its Cartesian coordinates $x(l,t)$ and $\eta(l,t)$, which satisfy the following constraint
\begin{align}
    x_l^2(l,t)+\eta_l^2(l,t) = S^2(t).\label{arclength eq}
\end{align}
Therefore, a more convenient way is to introduce the inclination angle $\theta(l,t)$ such that
\begin{align}
    \frac{x_l(l,t)}{S(t)} = \cos\theta(l,t), \qquad \frac{\eta_l(l,t)}{S(t)} = \sin\theta(l,t).\label{arclength eq2}
\end{align}
Eq. (\ref{arclength eq}) is then automatically satisfied and we can construct the interface from $\theta(l,t)$ and $S(t)$ by integrating (\ref{arclength eq2}). This formulation has benefit by reducing the number of unknowns from $O(3N)$ to $O(2N)$, where $N$ is the number of sample points on the interface.

\subsubsection{Equation for $S$}
Taking an infinitesimal interface element with length $\delta s$, the material derivative of $\delta s$ is
\begin{align}
    \frac{D\delta s}{Dt} = \big(\mathcal T_{i,s} - \theta_s \mathcal N_i\big)\delta s,\label{DdeltS}
\end{align}
where $\mathcal T_i$ denotes the tangential velocity of the interface, and $\mathcal N_i$ is the normal velocity (see Appendix \ref{appendix:Ddeltas}). Note that $\mathcal T_1 \neq \mathcal T_2$, but $\mathcal N_1 = \mathcal N_2$ from the kinematic boundary conditions. Therefore, we shall drop the subscript of $\mathcal N_i$ and use $\mathcal N$ hereafter. For periodic waves, we integrate Eq. (\ref{DdeltS}) and obtain the time-derivative of $S(t)$
\begin{align}
    \frac{\mathrm dS}{\mathrm d t} = \int_0^S\big(\mathcal T_{i,s} - \theta_s \mathcal N\big)\,\mathrm ds = -\int_0^1 \theta_l \mathcal N\,\mathrm dl,\label{dSdt}
\end{align}
where we have used the periodicity of $\mathcal T_i$.

\subsubsection{Function $\gamma$}
From (\ref{DdeltS}), we can also obtain the material derivative of $s$
\begin{align}
    \frac{D s}{D t} = \int_0^l \big(\mathcal T_{i,l} - \theta_l \mathcal N\big)\,\mathrm dl + \gamma(t),\label{DsDt}
\end{align}
where $\gamma(t)$ can be an arbitrary function. Note that $s$ and $t$ are independent variables, so $\partial s/\partial t \equiv 0$. On the other hand, $Ds/Dt \neq 0$ because for the same Lagrangian particle, its $s$ value changes with time. We can choose specific forms for $\gamma(t)$ and they have different physical meanings. A natural choice is to set $\gamma(t) \equiv 0$, which is equivalent to say $Ds/Dt = 0$ at $l=0$. The physical meaning is that we always fix $l=0$ onto the leftmost fluid particle. Without loss of generality, we let $i=1$ in (\ref{DsDt}) hereafter. Another computationally convenient choice is to set $l = 0$ onto the left boundary of the computational domain. One can then show that $\gamma(t) = \mathcal T_i(0,t)-\mathcal N(0,t)\tan \big(\theta(0,t)\big)$ (see Appendix \ref{appendix:gamma}). Hereafter, we shall denote the two choices by \textbf{Case I} and \textbf{Case II}
\begin{align}
    \gamma(t) = 
    \begin{cases}
        0, \qquad & \textbf{Case I}\\
        \mathcal T_1(0,t)-\mathcal N(0,t)\tan \big(\theta(0,t)\big), \qquad & \textbf{Case II}
    \end{cases}\label{deta0}
\end{align}
Note that \textbf{Case I} belongs to the mixed Eulerian-Lagrangian description and \textbf{Case II} is an Eulerian description.

\subsubsection{Equation for $\theta$}
The material derivative of $\theta$ satisfies
\begin{align}
    \frac{D \theta}{D t} = \mathcal N_s + \theta_s \mathcal T_i.\label{DthetaDs}
\end{align}
The derivation is in Appendix \ref{appendix:Ddeltas}. Expressing the material derivative in terms of $t$ and $l$, we obtain
\begin{align}
    \frac{D}{D t} = \frac{\partial}{\partial t} + \frac{D l}{D t} \frac{\partial}{\partial l}.
\end{align}
Since $l = s/S(t)$, we have
\begin{align}
    \frac{D l}{D t} = \frac{1}{S(t)}\frac{D s}{D t} - \frac{l}{S(t)}\frac{\mathrm d S}{\mathrm dt}.\label{dldt}
\end{align}
Substituting (\ref{dSdt}) and (\ref{DsDt}), we get the equation for $\theta$
\begin{align}
    \theta_t = \frac{\mathcal T_1(0,t)}{S}\theta_l + \frac{\mathcal N_l}{S} + \frac{\theta_l}{S}\bigg( \int_0^l \theta_l \mathcal N\,\mathrm dl - l\int_0^1 \theta_l \mathcal N\,\mathrm dl \bigg) - \frac{\gamma}{S}\theta_l\label{dtheta}
\end{align}

\subsubsection{Equation for $\varphi$}\label{section for phi}
In the Bernoulli equation (\ref{dynamic2}), the velocity potentials $\phi_1$ and $\phi_2$ are coupled together. Therefore, we introduce a density-weighted potential function $\varphi$
\begin{align}
    \varphi:= \varphi_1 - R\varphi_2 = \phi_1\Big(x(l,t),\eta(l,t),t\Big) - R\phi_2\Big(x(l,t),\eta(l,t),t\Big).
\end{align}
Note that $\varphi$ denotes the value of $\phi_1-R\phi_2$ evaluated on the interface. Taking the $s$-derivative, we get the density-weighted tangential velocity $\mathcal T$
\begin{align}
    \mathcal T:= \mathcal T_1 - R\mathcal T_2
\end{align}
The material derivative of $\varphi_i$ is
\begin{align}
    \frac{D\varphi_i}{Dt} = \phi_{i,t} + |\nabla \phi_i|^2 = \phi_{i,t} + \mathcal T_i^2 + \mathcal N^2.\label{dvarphi}
\end{align}
On the other hand, we have
\begin{align}
    \frac{D\varphi_i}{Dt} = \varphi_{i,t} + \frac{D l}{D t}\varphi_{i,l}.
\end{align}
Combining these and the Bernoulli equation, we can get the equations for $\varphi_1$ and $\varphi_2$
\begin{align}
    \varphi_{1,t} &= \mathcal T_1(0,t)\mathcal T_1 + \frac{1}{2}(\mathcal N^2-\mathcal T_1^2) + \mathcal T_1 \bigg( \int_0^l \theta_l \mathcal N\,\mathrm dl - l\int_0^1 \theta_l \mathcal N\,\mathrm dl \bigg) - \eta - p_1 - \mathcal T_1\gamma
    \label{dvarphi1},\\
    \varphi_{2,t} &= \mathcal T_1(0,t)\mathcal T_2 + \frac{1}{2}(\mathcal N^2-\mathcal T_2^2) + \mathcal T_2 \bigg( \int_0^l \theta_l \mathcal N\,\mathrm dl - l\int_0^1 \theta_l \mathcal N\,\mathrm dl \bigg) - \eta - \frac{p_2}{R} - \mathcal T_2\gamma\label{dvarphi2}.
\end{align}
Using the Young-Laplace condition
\begin{align}
    p_1 + \theta_s = p_2
\end{align}
to eliminate pressure $p_i$, we obtain the equation for $\varphi$
\begin{align}
    \varphi_t = \mathcal T_1(0,t)\mathcal T &- R \mathcal T_2(0,t)\mathcal T_2 + \frac{1}{2}(1-R)\mathcal N^2 - \frac12(\mathcal T_1^2 - R\mathcal T_2^2)\nonumber\\
    &+ \mathcal T \bigg( \int_0^l \theta_l \mathcal N\,\mathrm dl - l\int_0^1 \theta_l \mathcal N\,\mathrm dl \bigg) - (1-R)\eta + \frac{\theta_l}{S} - \mathcal T\gamma\label{dvarphi}
\end{align}

\subsubsection{Equations for $x_0$ and $\eta_0$}
The interface can be constructed by integrating (\ref{arclength eq2})
\begin{align}
    x &= S\int_0^l \cos\theta\,\mathrm dl + x_0,\label{x}\\
    \eta &= S\int_0^l \sin\theta\,\mathrm dl + \eta_0.\label{eta}
\end{align}
Depending on the choice of function $\gamma(t)$, the equations for $x_0$ and $\eta_0$ are
\begin{align}
    \frac{\mathrm dx_0}{\mathrm dt} = 
    \begin{cases}
        \mathcal T(0,t)\cos\big(\theta(0,t)\big) - \mathcal N(0,t)\sin\big(\theta(0,t)\big), \qquad & \textbf{Case I}\\
        0, \qquad & \textbf{Case II}
    \end{cases}\label{dx0}
\end{align}
\begin{align}
    \frac{\mathrm d\eta_0}{\mathrm dt} = 
    \begin{cases}
        \mathcal T(0,t)\sin\big(\theta(0,t)\big) + \mathcal N(0,t)\cos\big(\theta(0,t)\big), \qquad & \textbf{Case I}\\
        \mathcal N(0,t)/\cos\big(\theta(0,t)\big), \qquad & \textbf{Case II}
    \end{cases}\label{deta0}
\end{align}
It should be pointed out that the \textbf{Case II} choice is not always safe due to the possibility of $\theta = \pm \pi/2$ when waves become overhanging. Therefore, we prefer using \textbf{Case I} when there exist a potential of overturned waves. In the simulations, we expect that volume conservation is satisfied for all time
\begin{align}
    S\int_0^1 \eta \cos\theta\,\mathrm dl = 0.
\end{align}
This provides another way to determine $\eta_0$ without involving (\ref{deta0}) in computations.

\subsubsection{Equation for $\mathcal N$}
The mapping from $\varphi$ to $\mathcal N$ is the so called Dirichlet-to-Neumann operator in the theory of water waves. In our algorithm, we establish this relationship from boundary integral equations. For periodic waves with wave number $k$, we introduce the following complex mapping
\begin{align}
    \zeta = \mathrm e^{-\mathrm ikz},
\end{align}
which maps the physical region in a single spatial period to an annular region on the $\zeta$-plane. Since the complex velocity $w_{i} = u_{i}-\mathrm i v_{i}$ is an analytic function of $z=x+\mathrm iy$, it satisfies the Cauchy integral formula
\begin{align}
    w_{i}(\zeta_0) = \frac{1}{\mathrm i\pi}\oint_{C_{i}} \frac{w_{i}(\zeta)}{\zeta-\zeta_0}\,\mathrm d\zeta,\label{Cauchy integral}
\end{align}
where $C_{i} (i=1,2)$ denote boundaries of the lower and upper fluids on the new plane. Note that the complex velocity can be written in terms of $\mathcal T_i$, $\mathcal N$, and $\theta$ as
\begin{align}
    w_i = (\mathcal T_i - \mathrm i\mathcal N) \mathrm e^{-\mathrm i\theta}.
\end{align}
Substituting it into the Cauchy integral formula and taking the real and imaginary parts, we have the following four equations
\begin{align}
     \mathcal N(s_0) &= \int_0^S\Big( \mathcal A(s_0,s)-\mathcal B(s_0,s)\Big)\mathcal T_1(s)\,\mathrm ds + \int_0^S\Big( \mathcal C(s_0,s)+\mathcal D(s_0,s)\Big)\mathcal N(s)\,\mathrm ds,\label{integralforN}
     \\
     \mathcal N(s_0) &= \int_0^S\Big( \mathcal B(s_0,s)-\mathcal E(s_0,s)\Big)\mathcal T_2(s)\,\mathrm ds - \int_0^S\Big( \mathcal D(s_0,s)+\mathcal F(s_0,s)\Big)\mathcal N(s)\,\mathrm ds,\label{integralforN2}\\
     \mathcal T_1(s_0) &= -\int_0^S\Big( \mathcal C(s_0,s)-\mathcal D(s_0,s)\Big)\mathcal T_1(s)\,\mathrm ds + \int_0^S\Big( \mathcal A(s_0,s)+\mathcal B(s_0,s)\Big)\mathcal N(s)\,\mathrm ds,\label{integralforT1}\\
    \mathcal T_2(s_0) &= -\int_0^S\Big( \mathcal D(s_0,s)-\mathcal F(s_0,s)\Big)\mathcal T_2(s)\,\mathrm ds - \int_0^S\Big( \mathcal B(s_0,s)+\mathcal E(s_0,s)\Big)\mathcal N(s)\,\mathrm ds.\label{integralforT2}
\end{align}
The functions $\mathcal A(s_0,s)$ to $\mathcal F(s_0,s)$ are
\begin{align}
    \mathcal A(s_0,s) = \frac k\pi\text{Im}\bigg( \frac{\mathrm e^{\mathrm i\theta(s_0)}\mathrm e^{-2kh_1}/\zeta^*(s)}{\mathrm e^{-2kh_1}/\zeta^*(s)-\zeta(s_0)}\bigg)&,\qquad \mathcal B(s_0,s) = \frac k\pi\text{Im}\bigg(\frac{\mathrm e^{\mathrm i\theta(s_0)}\zeta(s)}{\zeta(s)-\zeta(s_0)} \bigg),\\
    \mathcal C(s_0,s) = \frac k\pi\text{Re}\bigg( \frac{\mathrm e^{\mathrm i\theta(s_0)}\mathrm e^{-2kh_1}/\zeta^*(s)}{\mathrm e^{-2kh_1}/\zeta^*(s)-\zeta(s_0)}\bigg)&,\qquad \mathcal D(s_0,s) = \frac k\pi\text{Re}\bigg(\frac{\mathrm e^{\mathrm i\theta(s_0)}\zeta(s)}{\zeta(s)-\zeta(s_0)} \bigg),\\
    \mathcal E(s_0,s) = \frac k\pi\text{Im}\bigg(\frac{\mathrm e^{\mathrm i\theta(s_0)}/\zeta^*(s)}{1/\zeta^*(s)-\zeta(s_0)/\mathrm e^{2kh_2}} \bigg)&,\qquad \mathcal F(s_0,s) = \frac k\pi\text{Re}\bigg(\frac{\mathrm e^{\mathrm i\theta(s_0)}/\zeta^*(s)}{1/\zeta^*(s)-\zeta(s_0)/\mathrm e^{2kh_2}} \bigg),\label{A2F}
\end{align}
where Re and Im denote the real and imaginary parts, and $\zeta^*(s)$ is the complex conjugate of $\zeta(s)$. When $s\rightarrow s_0$
\begin{align*}
    \frac{\mathrm e^{\mathrm i\theta(s_0)}\zeta(s)}{\zeta(s)-\zeta(s_0)} \rightarrow \frac{\mathrm i}{k (s-s_0)} + \frac{\mathrm e^{\mathrm i\theta(s_0)}}{2} + \frac{\theta_s(s_0)}{2k},
\end{align*}
thus function $\mathcal B(s_0,s)$ has a removable singularity. When evaluating,  say the second integral in (\ref{integralforN}), $\mathcal B(s_0,s)\mathcal T_1(s)$ can be replaced by $\mathcal T_{1,s}(s_0)/\pi$ at $s=s_0$. Note that for given $\mathcal T$ and $z$, Eqs. (\ref{integralforN})-(\ref{integralforT2}) actually contain two unknowns, i.e. $\mathcal N$ and one of $\mathcal T_i, (i=1,2)$ (the other can be calculated from $\mathcal T$). 
For convenience, we introduce a new unknown $\mathcal U := \mathcal T_1+\mathcal T_2$, then we have
\begin{align}
    \mathcal T_1 = \frac{\mathcal T + R\mathcal U}{1+R} , \qquad \mathcal T_2 = \frac{\mathcal U-\mathcal T}{1+R}.\label{T12}
\end{align}
To get the equation for $\mathcal N$ and $\mathcal U$, we add the last two integral equations and discretize the integral by using the trapezoid rule
\begin{align}
    \mathcal U_{m} = \sum_{n=1}^{N}(\mathcal D_{mn}-\mathcal C_{mn})\mathcal T_{1,n}\Delta s + \sum_{n=1}^{N}(\mathcal F_{mn}-\mathcal D_{mn})\mathcal T_{2,n}\Delta s + \sum_{n=1}^{N}(\mathcal A_{mn}-\mathcal E_{mn})\mathcal N_{n}\Delta s,
\end{align}
where $\mathcal T_{i,m} = \mathcal T_i(s_m)$, $\mathcal D_{mn} = \mathcal D(s_m,s_n)$, etc.
Multiplying the second integral equation by $R$ and then adding to the first one, we have
\begin{align}
    (1+R)\mathcal N_m = \sum_{n=1}^{N}\mathcal A_{mn}\mathcal T_{1,n}\Delta s - R\sum_{n=1}^{N}\mathcal E_{mn}\mathcal T_{2,n}\Delta s &+ \sum_{n=1}^{N}\Big(\mathcal C_{mn}-R\mathcal F_{mn} + (1-R)\mathcal D_{mn}\Big)\mathcal N_{n}\Delta s\nonumber\\
    &-\underbrace{\sum_{n=1, n\neq m}^{N}\mathcal B_{mn}\mathcal T_{n}\Delta s - \frac1{\pi}\frac{\partial\mathcal T_{n}}{\partial s}\Delta s}_{\text{known}}.
\end{align}
Substituting (\ref{T12}), we have
\begin{align}
    \mathcal U_m =& \sum_{n=1}^{N}\frac{\mathcal F_{mn}-R\mathcal C_{mn}-(1-R)\mathcal D_{mn}}{1+R}\mathcal U_{n}\Delta s + \sum_{n=1}^{N}(\mathcal A_{mn}-\mathcal E_{mn})\mathcal N_{n}\Delta s\nonumber\\
    &+ \underbrace{\sum_{n=1}^{N}\frac{2\mathcal D_{mn}-\mathcal C_{mn}-\mathcal F_{mn}}{1+R}\mathcal T_{n}\Delta s}_{\text{known}},\\
    \mathcal N_{m} =& \sum_{n=1}^{N}\frac{R(\mathcal A_{mn}-\mathcal E_{mn})}{(1+R)^2}\mathcal U_{n}\Delta s + \sum_{n=1}^{N}\frac{ \mathcal C_{mn} - R \mathcal F_{mn} + (1-R)\mathcal D_{mn}}{1+R} \mathcal N_{n}\Delta s\nonumber\\
    &+ \underbrace{\sum_{n=1}^{N}\frac{\mathcal A_{mn} + R \mathcal E_{mn}}{(1+R)^2}\mathcal T_{n}\Delta s - \sum_{n=1, n\neq m}^{N}\frac{\mathcal B_{mn}}{1+R}\mathcal T_{n}\Delta s - \frac{1}{\pi(1+R)}\frac{\partial\mathcal T_n}{\partial s}\Delta s}_{\text{known}}.
\end{align}
They can be written into a more compact form
\begin{align}
    \begin{pmatrix}
    I - \frac{\mathcal F-R\mathcal C-(1-R)\mathcal D}{1+R}\Delta s & (\mathcal E - \mathcal A)\Delta s\\
    \frac{R(\mathcal E - \mathcal A)}{(1+R)^2}\Delta s & I - \frac{ \mathcal C - R \mathcal F + (1-R)\mathcal D}{1+R}\Delta s
     \end{pmatrix}\cdot
     \begin{pmatrix}
    \mathcal U\\
    \mathcal N
     \end{pmatrix}
     = 
     \begin{pmatrix}
    \frac{2\mathcal D-\mathcal C-\mathcal F}{1+R}\cdot\mathcal T\\
    \frac{\mathcal A + R \mathcal E}{(1+R)^2}\cdot\mathcal T - \frac{\mathcal B}{1+R}\cdot\mathcal T - \frac{\partial\mathcal T/\partial s}{\pi(1+R)}
     \end{pmatrix}\Delta s,\label{iteration2}
\end{align}
where $I$ and $0$ denote the identity matrix and the zero matrix. In deep-water case, i.e. $h_{1,2}\rightarrow\infty$, functions $\mathcal A$, $\mathcal C$, $\mathcal E$ and $\mathcal F$ vanish and (\ref{iteration2}) reduces to 
\begin{align}
    \begin{pmatrix}
    I + \frac{1-R}{1+R}\mathcal D\Delta s & 0\\
    0 & I - \frac{1-R}{1+R}\mathcal D\Delta s
     \end{pmatrix}\cdot
     \begin{pmatrix}
    \mathcal U\\
    \mathcal N
     \end{pmatrix}
     = 
     \begin{pmatrix}
    \frac{2}{1+R}\mathcal D\cdot\mathcal T\\
    - \frac{1}{1+R}\mathcal B\cdot\mathcal T - \frac{1}{\pi(1+R)}\frac{\partial\mathcal T}{\partial s}
     \end{pmatrix}\Delta s.
\end{align}

\subsection{Surface waves}
The preceding formulation can be employed to simulate surface waves by setting $R = 0$ in Eq. (\ref{dvarphi}) and using (\ref{dSdt}), (\ref{dtheta}), and (\ref{dx0}). Using the GMRES, the boundary integral (\ref{integralforN}) is used to find $\mathcal N$ for given $\mathcal T_1$ and $\theta$.

\subsection{Numerical implementation}
We discretize the interface(surface) by $N$ equally spaced grid points whose values of $l$ are
\begin{align}
    l_m = \frac{m-1}{N}, \qquad m = 1,2\cdots,N.
\end{align}
All spatial-derivatives with respect to $s$ can be calculated from their $l$-derivatives and they are computed by using the fft function in Matlab. The definite integrals in the boundary-integral equations have been discretized by using the trapezoid rule, which gives spectral accuracy for periodic functions. In the GMRES iterations, we set the threshold of convergence to be $10^{-12}$. To perform time integration, we apply the fourth-order Runge-Kutta method in the Fourier space instead of the physical space. To measure the numerical accuracy, we define the relative error of energy $E_r(t)$
\begin{align}
    E_r(t) := \frac{|E(t)-E(0)|}{E(0)},
\end{align}
where $E(t)$ is the energy calculated in one spatial period
\begin{align}
    E(t) = \frac{1}{2}\iint_{-h_1<y<\eta(x)} |\nabla \phi_1|^2\,\mathrm dy\mathrm dy &+ \frac{R}{2}\iint_{\eta(x)<y<h_2}|\nabla \phi_2|^2\,\mathrm dx\mathrm dy\nonumber\\
    &+ \frac{1-R}{2}\int_{0}^{2\pi/k}\eta^2\,\mathrm dx + \int_{0}^{2\pi/k} \big( \sqrt{1+\eta_x^2}-1 \big)\,\mathrm dx.
\end{align}
Using the divergence theorem, $E(t)$ can be reduced to integrals evaluated on the interface. In deep-water case, it reads
\begin{align}
    E(t) = -\frac{1}{2}\int_{0}^{S} \varphi\mathcal N\,\mathrm ds + \frac{1-R}{2}\int_{0}^{S}\eta^2\cos\theta\,\mathrm ds + \int_{0}^{S} \big( 1-\cos\theta \big)\,\mathrm ds.
\end{align}

\section{Travelling waves}
In this section, we briefly describe the numerical method used for interfacial travelling waves in deep-water case and their linear stability analysis. These solutions will be used as initial conditions for various unsteady simulations to study their stability.
\subsection{Formulation and numerical implementation}
It is computationally convenient to choose a moving frame of reference where travelling waves become time-independent. The kinematic boundary conditions (\ref{kinematic1}) and (\ref{kinematic2}) become
\begin{align}
    0 = \phi_{i,x}\eta_x - \phi_{i,y} = \sqrt{1+\eta_x^2}\,\mathcal N.
\end{align}
Therefore, $\mathcal N \equiv 0$ on the interface in the moving frame. The Bernoulli equation (\ref{dynamic2}) becomes
\begin{align}
    \frac{1}{2}\mathcal T_1^2 - \frac{R}{2}\mathcal T_2^2 + (1-R)S\bigg(\int_0^l\sin(\theta)\,\mathrm dl + \eta_0 \bigg) - \frac{\theta_l}{S} = B,\label{dynamic}
\end{align}
where $B$ denotes the unknown Bernoulli constant. The boundary integrals (\ref{integralforT1}) and (\ref{integralforT2}) become
\begin{align}
    \mathcal T_1(l_0)\sin\theta(l_0) &= -\frac{kS}{\pi}\int_{0}^{1}\text{Im}\bigg( \frac{1}{1-\zeta(l_0)/\zeta(l)} \bigg)\mathcal T_1(l)\,\mathrm dl,\label{integral1}\\
    \mathcal T_2(l_0)\sin\theta(l_0) &= \frac{kS}{\pi}\int_{0}^{1}\text{Im}\bigg( \frac{1}{1-\zeta(l_0)/\zeta(l)} \bigg)\mathcal T_2(l)\,\mathrm dl.\label{integral2}
\end{align}
We focus on symmetric solution, i.e. waves are invariant under reflection with respect to $y$-axis. Therefore, $\mathcal T_1$ and $\mathcal T_2$ are even functions of $x$ and $\theta$ is an odd function of $x$. We can write down their Fourier series with respect to $l$
\begin{align}
    \mathcal T_1 = \sum_{n=0}^{\infty}a_n\cos(2\pi n l),\quad
    \mathcal T_2 = \sum_{n=0}^{\infty}b_n\cos(2\pi n l),\quad
    \theta = \sum_{n=1}^{\infty}c_n\sin(2\pi n l).
\end{align}
Therefore, for $N+1$ equally spaced grid points $l_m$
\begin{align}
    l_m = \frac{m-1}{N}, \quad m = 1,2,\cdots,N+1
\end{align}
only those values of unknowns on the first $N/2+1$ points are necessary for computation. Truncating the Fourier series after $N/2$ terms gives $3N+2$ unknowns $a_0,\cdots,a_{N/2},b_0,\cdots,b_{N/2},c_1,\cdots,c_{N/2}$. Together with $\eta_0$, $B$, and $S$, we have $3N/2+5$ unknowns to solve. The Bernoulli equation (\ref{dynamic}) is satisfied on the first $N/2+1$ grid points. The boundary integrals (\ref{integral1}) and (\ref{integral2}) are evaluated on $N/2$ mid-points 
\begin{align}
    l^{\text{m}}_m = \frac{l_m+l_{m+1}}{2}, m = 1,2,\cdots,\frac{N}{2}
\end{align}
to remove the singularity. To make the system solvable, we need to impose four extra constraints
\begin{align}
    \int_{0}^{1}\eta \cos\theta\,\mathrm dl = 0,\qquad
    \int_{0}^{1} \cos\theta\,\mathrm dl = \frac{2\pi}{kS},\label{constraint}\\
    \int_{0}^{1}\mathcal T_1\,\mathrm dl = -\frac{2\pi c}{kS},\qquad
    \int_{0}^{1}\mathcal T_2\,\mathrm dl = -\frac{2\pi c}{kS}\label{wavespeed},
\end{align}
where $c$ denotes the given wave speed. The first and second equations guarantee volume conservation and periodicity in $x$-direction. The last two equations come from the irrotational condition
\begin{align}
    \oint \mathcal T_{i}\,\mathrm ds = \iint \nabla \times \vec u_{i}\,\mathrm dx \mathrm dy = 0.
\end{align}
In the moving frame of reference, $\mathcal T_{i} \rightarrow -c$ when $|y| \rightarrow \infty$, we have
\begin{align}
    \oint \mathcal T_1\,\mathrm ds =& -\frac{2\pi c}{k} - S\int_{0}^{1}\mathcal T_1\,\mathrm dl,\\
    \oint \mathcal T_2\,\mathrm ds =& \frac{2\pi c}{k} + S\int_{0}^{1}\mathcal T_2\,\mathrm dl.
\end{align}
Substututing the Fourier series of $\mathcal T_{i}$, (\ref{wavespeed}) become
\begin{align}
    a_0 = -\frac{2\pi c}{kS} = b_0.
\end{align}

\subsection{Newton-Krylov method}\label{NewtonKyrlov}
A traditional but effective way to solve nonlinear equations is Newton's method, which has been widely used in the water wave community. A drawback of this method is that one has to update the Jacobian matrix per iteration. For complex nonlinear systems involving boundary integrals or non-analytical operations, this is usually conducted by using finite difference method to approximate the derivatives. However, the time cost increases violently with $N$, making it difficult for large-scale computations. In the recent decades, Newton-Krylov method becomes popular in the field of computational physics and has been employed to water waves \cite{pethiyagoda2014jacobian, ctugulan2022three}. It does not require to construct the Jacobian matrix, thus has much smaller time cost than Newton's method. In this subsection, we briefly explain the algorithm. Interested readers are referred to \cite{knoll2004jacobian} and the literature therein for more technique details. 

For $N$ nonlinear equations
\begin{align}
    \boldsymbol F(\boldsymbol x) = \boldsymbol 0,
\end{align}
where $\boldsymbol x$ is a $N$-dimensional unknown, Newton's method requires an initial guess $\boldsymbol x_0$ and updates it
\begin{align}
    \boldsymbol x_{n+1} = \boldsymbol x_{n} + \delta \boldsymbol x_n
\end{align}
by solving the following linear equation
\begin{align}
    \boldsymbol J_n(\boldsymbol x_n) \delta \boldsymbol x_n = - \boldsymbol F(\boldsymbol x_n),\label{newton}
\end{align}
where $\boldsymbol J_n(\boldsymbol x_n) = \partial \boldsymbol F(\boldsymbol x_n)/\partial \boldsymbol x_n$ is the Jacobian matrix. The genius of Newton-Krylov method is reflected in the following aspects
\begin{itemize}
    \item One does not need to solve (\ref{newton}) exactly because $\delta \boldsymbol x_n$ only provides an approximate direction to update $\boldsymbol x_n$. This is the underlying idea of the so-called ``inexact Newton's method".
    \item One can search for the approximate solution of (\ref{newton}) from its Krylov subspace $\mathcal K_r$. A Krylov subspace corresponding to a linear equation $\boldsymbol A \boldsymbol x = \boldsymbol b$ is defined as
    \begin{align}
        \mathcal K_r = \text{span}\{\boldsymbol b, \boldsymbol A \boldsymbol b, \boldsymbol A^2 \boldsymbol b, \cdots, \boldsymbol A^r \boldsymbol b\}.
    \end{align}
    Usually one can construct a well approximate solution when $r\ll N$. By comparison, directly solving (\ref{newton}) is equivalent to search for solution in the $N$ dimensional vector space with standard basis $\{\boldsymbol{\mathrm{e}}_1, \boldsymbol{\mathrm{e}}_2, \cdots, \boldsymbol{\mathrm{e}}_N\}$. To construct the Krylov subspace, one only needs to calculate the matrix-vector product. From definition,
    \begin{align}
        \boldsymbol J_n(\boldsymbol x_n)\delta \boldsymbol x_n \approx \frac{\boldsymbol F(\boldsymbol x_n + \epsilon \delta \boldsymbol x_n) - \boldsymbol F(\boldsymbol x_n)}{\epsilon},
    \end{align}
    where $\epsilon\ll 1$ is a small constant. Therefore, one can build the Krylov subspace without calculating the Jacobian matrix $\boldsymbol J_n(\boldsymbol x_n)$.
\end{itemize}
It is worth mentioning that the Krylov space is usually constructed using the ``Arnoldi iteration" instead of the successive power method \cite{trefethen2022numerical}. This is a key step of the GMRES. Additionally, one needs a preconditioning matrix $\boldsymbol P$ to solve (\ref{newton}) effectively. $\boldsymbol P$ is an approximation of the Jacobian matrix $\boldsymbol J_n(\boldsymbol x_n)$ and is used to cluster its eigenvalues. In our computation, the part of the Jacobian matrix corresponding to Eq. (\ref{dynamic}) and (\ref{constraint}) are calculated analytically and used. For boundary integral equations (\ref{integral1}) and (\ref{integral2}), only the term on the left hand side is used in $\boldsymbol P$.

In a nutshell, the Newton-Krylov method consists of an outer iteration to update $\boldsymbol x_n$ and an inner iteration to obtain correction vector $\delta \boldsymbol x_n$ approximately. In our computation, we set a maximum value $N_{max}$ to restrict the number of inner iteration. When it terminates, $\delta \boldsymbol x_n$ is passed to the outer iteration to check whether the convergence condition, $\lVert \boldsymbol F(\boldsymbol x_n)\rVert_{\infty}<10^{-11}$, is satisfied. Based on our numerical experiments, $N_{max}$ is usually set to be $10$. When solutions become highly nonlinear, $N_{max}$ needs to be increased gradually but without exceeding $40$. Increasing the value of $N_{max}$ too much can cause failure of convergence. 

\subsection{Linear stability}
In the moving frame of reference, we consider the following perturbations superposed on the steady solutions $\eta(x)$, $\Phi_i(x,y)$ and $\Psi_i(x,y) (i=1,2)$, where $\Psi_i$ are the stream functions in each layer
\begin{align}
    \eta(x,t) \rightarrow& \eta(x) + \mathrm e^{\lambda t}\tilde \eta(x),\\
    \phi_1(x,y,t) \rightarrow \Phi_1(x,y) + \mathrm e^{\lambda t}\tilde \phi_1(x,y),&\quad
    \psi_1(x,y,t) \rightarrow \Psi_1(x,y) + \mathrm e^{\lambda t}\tilde \psi_1(x,y),\\
    \phi_2(x,y,t) \rightarrow \Phi_2(x,y) + \mathrm e^{\lambda t}\tilde \phi_2(x,y),&\quad
    \psi_2(x,y,t) \rightarrow \Psi_2(x,y) + \mathrm e^{\lambda t}\tilde \psi_2(x,y),
\end{align}
where the tiled terms represent small perturbations.

\subsubsection{Formulation from Calvo \& Akylas}
Following the work of \cite{calvo2002stability,calvo2003interfacial}, we have the following equations for $\tilde\eta$, $\tilde\phi_1$ and $\tilde\phi_2$ after linearizing Eqs. (\ref{kinematic1}), (\ref{kinematic2}), and (\ref{dynamic2})
\begin{align}
    \lambda\Tilde{\eta} =& -\frac{1}{\cos\theta}\frac{\mathrm d\mathcal P_1(\Tilde{\phi_1})}{\mathrm ds} - \frac{1}{\cos\theta}\frac{\mathrm d}{\mathrm ds}\big(\mathcal T_1\cos\theta\big)\tilde\eta - \mathcal T_1\frac{\mathrm d\tilde\eta}{\mathrm ds},\label{kinematic_interfacial1}\\
    \lambda\Tilde{\eta} =& -\frac{1}{\cos\theta}\frac{\mathrm d\mathcal P_2(\Tilde{\phi_2})}{\mathrm ds} - \frac{1}{\cos\theta}\frac{\mathrm d}{\mathrm ds}\big(\mathcal T_2\cos\theta\big)\tilde\eta - \mathcal T_2\frac{\mathrm d\tilde\eta}{\mathrm ds},\label{kinematic_interfacial2}\\
    \lambda(\Tilde{\phi
    }_1 - R\tilde\phi_2)=& -\mathcal T_1\frac{\mathrm d\Tilde \phi_1}{\mathrm ds} + R\mathcal T_2\frac{\mathrm d\Tilde \phi_2}{\mathrm ds} -\bigg( \mathcal T_1\frac{\mathrm d (\mathcal T_1\sin\theta)}{\mathrm ds} - R\mathcal T_2\frac{\mathrm d (\mathcal T_2\sin\theta)}{\mathrm ds} + 1 -R\bigg)\Tilde\eta\nonumber\\
    &+ \frac{1}{\cos\theta}\frac{\mathrm d}{\mathrm ds}\bigg(\cos^2\theta\frac{\mathrm d\tilde\eta}{\mathrm ds}\bigg),\label{bernouli_interfacial}
\end{align}
where $\mathcal P_1$ and $\mathcal P_2$ are operators relating $\tilde \phi_{1,2}$ and $\tilde\psi_{1,2}$ and can be found from Cauchy's integral formula (\ref{Cauchy integral}). This leads to a generalised eigenvalue problem
\begin{align}
    \begin{pmatrix}
        M_1 & M_2 & 0\\
        M_3 & 0 & M_4\\
        M_5 & M_6 & M_7
    \end{pmatrix}
    \begin{pmatrix}
        \tilde\eta\\
        \tilde\phi_1\\
        \tilde\phi_2
    \end{pmatrix}
    = \lambda
    \begin{pmatrix}
        I & 0 & 0\\
        I & 0 & 0\\
        0 & I & -R I
    \end{pmatrix}
    \begin{pmatrix}
        \tilde\eta\\
        \tilde\phi_1\\
        \tilde\phi_2
    \end{pmatrix}\label{eigequation}
\end{align}
where $0$ and $I$ represent the zero and identity matrices. Although it can be solved by the built-in Matlab function eig, we found this is rather time-consuming and numerically sensitive. Therefore, we derive a new formulation to perform the linear stability analysis.

\subsubsection{New formulation}
Motivated by the time-stepping algorithm, we introduce two variables $\tilde\xi$ and $\tilde\chi$
\begin{align}
    \tilde\xi := \tilde\phi_1 - R\tilde\phi_2, \qquad \tilde\chi := \tilde\phi_1 + \tilde\phi_2
\end{align}
$\tilde \phi_1$ and $\tilde\phi_2$ can be obtained from the following two relations
\begin{align}
    \tilde \phi_1 = \frac{\tilde\xi + R\tilde\chi}{1+R}, \qquad \tilde \phi_2 = \frac{\tilde\chi - \tilde\xi}{1+R},\label{phi12}
\end{align}
Repacing $w_i$ by $\tilde\phi_i+\mathrm i\tilde\psi_i$ in (\ref{Cauchy integral}), we have
\begin{align}
    \tilde\phi_1 =& -\mathcal M_r \tilde \psi_1 - \mathcal M_i \tilde \phi_1,\\
    \tilde\phi_2 =& \mathcal M_r \tilde \psi_2 + \mathcal M_i \tilde \phi_2,
\end{align}
where $\mathcal M_r$ and $\mathcal M_i$ represent the real and imaginary part of the boundary-integral operator. Adding them together, we have
\begin{align}
    \tilde\chi = -\mathcal M_r (\tilde\psi_1 - \tilde\psi_2) - \mathcal M_i \tilde\phi_1 + \mathcal M_i \tilde\phi_2.\label{chi}
\end{align}
The following identity is implied by subtracting (\ref{kinematic_interfacial1}) and (\ref{kinematic_interfacial2})
\begin{align}
    \tilde \psi_1 - \tilde\psi_2 = \tilde\eta(\mathcal T_2-\mathcal T_1)\cos\theta.
\end{align}
Thus (\ref{chi}) becomes
\begin{align}
    \tilde\chi = -\mathcal M_r \bigg( \tilde\eta(\mathcal T_2-\mathcal T_1)\cos\theta \bigg) - \mathcal M_i \tilde\phi_1 + \mathcal M_i \tilde\phi_2.
\end{align}
Replacing $\tilde\phi_1$ and $\tilde\phi_2$ by $\tilde\xi$ and $\tilde\chi$, we obtain
\begin{align}
    \bigg( I + \frac{(R-1)\mathcal M_i}{1+R} \bigg)\tilde\chi = -\mathcal M_r \bigg( \tilde\eta(\mathcal T_2-\mathcal T_1)\cos\theta \bigg) - \frac{2\mathcal M_i}{1+R}\tilde\xi,
\end{align}
or in a more compact form
\begin{align}
    \tilde\chi = \mathcal Q_1 \tilde \eta + \mathcal Q_2 \tilde \xi.
\end{align}
Now we can rewrite the kinematic boundary condition and dynamic boundary condition
\begin{itemize}
    \item For Eq. (\ref{kinematic_interfacial1})
    \begin{align}
        \lambda \tilde\eta =& -\frac{1}{\cos\theta}\frac{\mathrm d}{\mathrm ds}\Big(\mathcal P_1(\tilde\phi_1) + \tilde\eta \mathcal T_1\cos\theta\Big) = -\frac{1}{\cos\theta}\frac{\mathrm d}{\mathrm ds}\bigg(\frac{\mathcal P_1 \tilde\xi + R \mathcal P_1 \tilde\chi}{1+R} + \tilde\eta \mathcal T_1\cos\theta \bigg)\nonumber\\
        =& -\frac{1}{\cos\theta}\frac{1}{1+R}\frac{\mathrm d}{\mathrm ds}\big( \mathcal P_1\tilde\xi \big) - \frac{1}{\cos\theta}\frac{R}{1+R}\frac{\mathrm d}{\mathrm ds}\big( \mathcal P_1\mathcal Q_1\tilde\eta + \mathcal P_1\mathcal Q_2\tilde\xi \big) - \frac{1}{\cos\theta}\frac{\mathrm d}{\mathrm ds}\big( \tilde\eta \mathcal T_1\cos\theta \big).
    \end{align}
    \item For Eq. (\ref{bernouli_interfacial}), the kinetic energy term becomes
    \begin{align}
        -\mathcal T_1\frac{\mathrm d\tilde\phi_1}{\mathrm ds} + R \mathcal T_2\frac{\mathrm d\tilde\phi_2}{\mathrm ds} =& -\mathcal T_1\frac{\mathrm d}{\mathrm ds}\bigg( \frac{\tilde\xi + R\tilde\chi}{1+R} \bigg) + R \mathcal T_2\frac{\mathrm d}{\mathrm ds}\bigg( \frac{\tilde\chi - \tilde\xi}{1+R} \bigg)\nonumber\\
        =& -\frac{\mathcal T_1+R\mathcal T_2}{1+R}\frac{\mathrm d\tilde\xi}{\mathrm ds} - \frac{R(\mathcal T_1-\mathcal T_2)}{1+R}\frac{\mathrm d}{\mathrm ds}\big( \mathcal Q_1\tilde\eta + \mathcal Q_2\tilde\xi \big).
    \end{align}
    So (\ref{bernouli_interfacial}) becomes
    \begin{align}
        \lambda \tilde\xi =& -\frac{\mathcal T_1+R\mathcal T_2}{1+R}\frac{\mathrm d\tilde\xi}{\mathrm ds} - \frac{R(\mathcal T_1-\mathcal T_2)}{1+R}\frac{\mathrm d}{\mathrm ds}\big( \mathcal Q_1\tilde\eta + \mathcal Q_2\tilde\xi \big),\nonumber\\
        &-\bigg( \mathcal T_1\frac{\mathrm d (\mathcal T_1\sin\theta)}{\mathrm ds} - R \mathcal T_2\frac{\mathrm d (\mathcal T_2\sin\theta)}{\mathrm ds} + 1-R \bigg)\tilde\eta
       + \frac{1}{\cos\theta}\frac{\mathrm d}{\mathrm ds}\bigg(\cos^2\theta\frac{\mathrm d\tilde\eta}{\mathrm ds}\bigg).
    \end{align}
\end{itemize}
Ultimately, we have
\begin{align}
    \lambda \tilde\eta =& -\frac{1}{\cos\theta}\frac{\mathrm d}{\mathrm ds}\bigg( \frac{\mathcal P_1 + R \mathcal P_1 \mathcal Q_2}{1+R} \tilde\xi\bigg) - \frac{1}{\cos\theta}\frac{\mathrm d}{\mathrm ds}\bigg( \tilde\eta \mathcal T_1\cos\theta \bigg)- \frac{1}{\cos\theta}\frac{\mathrm d}{\mathrm ds}\bigg( \frac{R\mathcal P_1\mathcal Q_1\tilde\eta}{1+R}\bigg)\label{newkinematic}\\
    \lambda \tilde\xi =& -\frac{\mathcal T_1+R\mathcal T_2}{1+R}\frac{\mathrm d\tilde\xi}{\mathrm ds} - \frac{R(\mathcal T_1-\mathcal T_2)}{1+R}\frac{\mathrm d\mathcal Q_2(\tilde\xi)}{\mathrm ds}-\bigg( \mathcal T_1\frac{\mathrm d (\mathcal T_1\sin\theta)}{\mathrm ds} - R \mathcal T_2\frac{\mathrm d (\mathcal T_2\sin\theta)}{\mathrm ds} + 1-R \bigg)\tilde\eta\nonumber
    \\
    &+ \frac{1}{\cos\theta}\frac{\mathrm d}{\mathrm ds}\bigg(\cos^2\theta\frac{\mathrm d\tilde\eta}{\mathrm ds}\bigg)- \frac{R(q_1-q_2)}{1+R}\frac{\mathrm d\mathcal Q_1(\tilde\eta)}{\mathrm ds}\label{newbernoulli}
\end{align}
This is an eigenvalue problem of standard form. When $R = 0$, it becomes the formulation in \cite{tanaka1986stability,calvo2002stability}. Compared with Eqs. (\ref{kinematic_interfacial1})-(\ref{bernouli_interfacial}), our formulation reduces the number of unknowns by a third and turns out to be more stable numerically.

\section{Numerical results}
In this section, we firstly show some numerical simulations of surface waves, including propagation and breaking of periodic waves, head-on collision of solitary waves. Then we move to the simulations of interfacial waves, focus on the stability of interfacial gravity-capillary solitary waves, and compare the results with those from linear stability analysis. Finally, we display a simulation of vortex roll-up due to the K-H instability.
\subsection{Surface gravity waves}
\subsubsection{Propagation of periodic waves}
\begin{figure}[!h]
    \centering
    \subfigure[]{
    \begin{minipage}{0.48\textwidth}
    \centering\includegraphics[width=\textwidth]{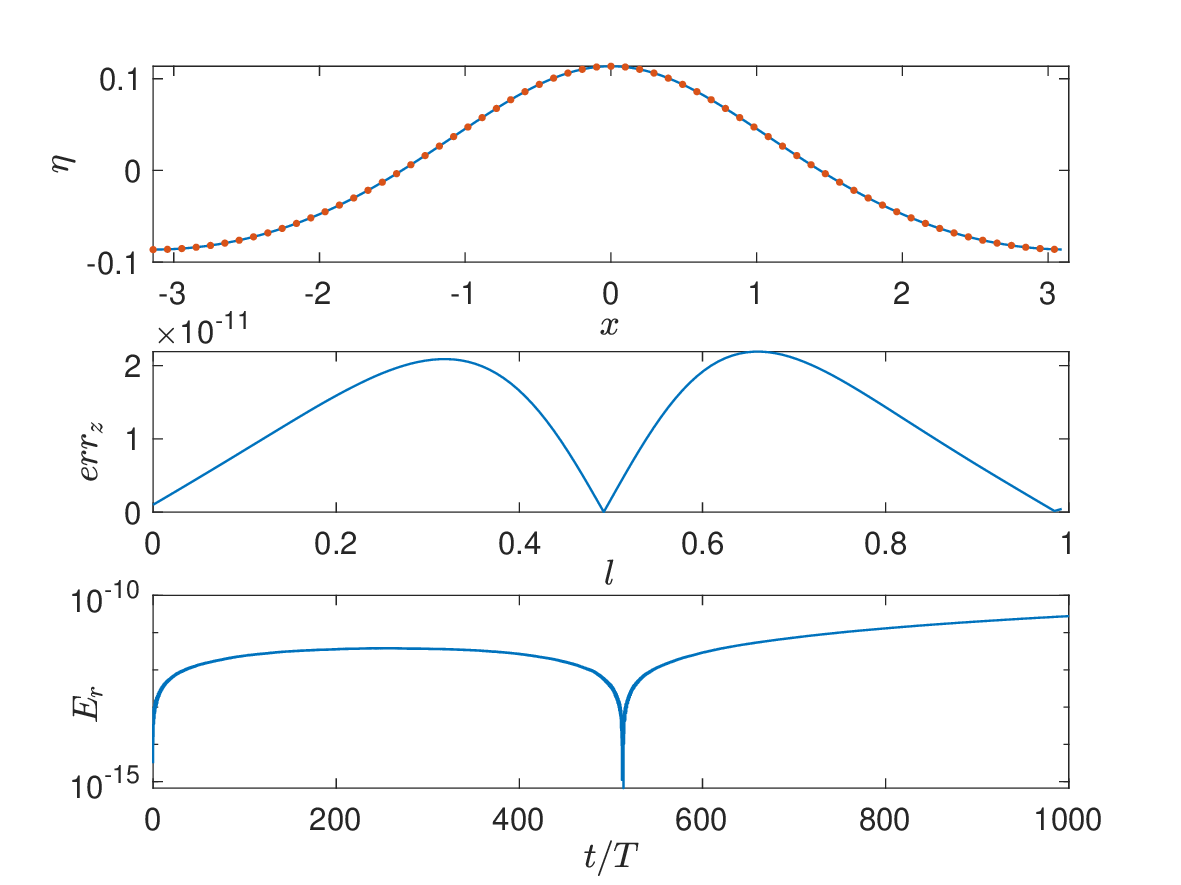}
    \end{minipage}
    }
    \subfigure[]{
    \begin{minipage}{0.48\textwidth}
    \centering\includegraphics[width=\textwidth]{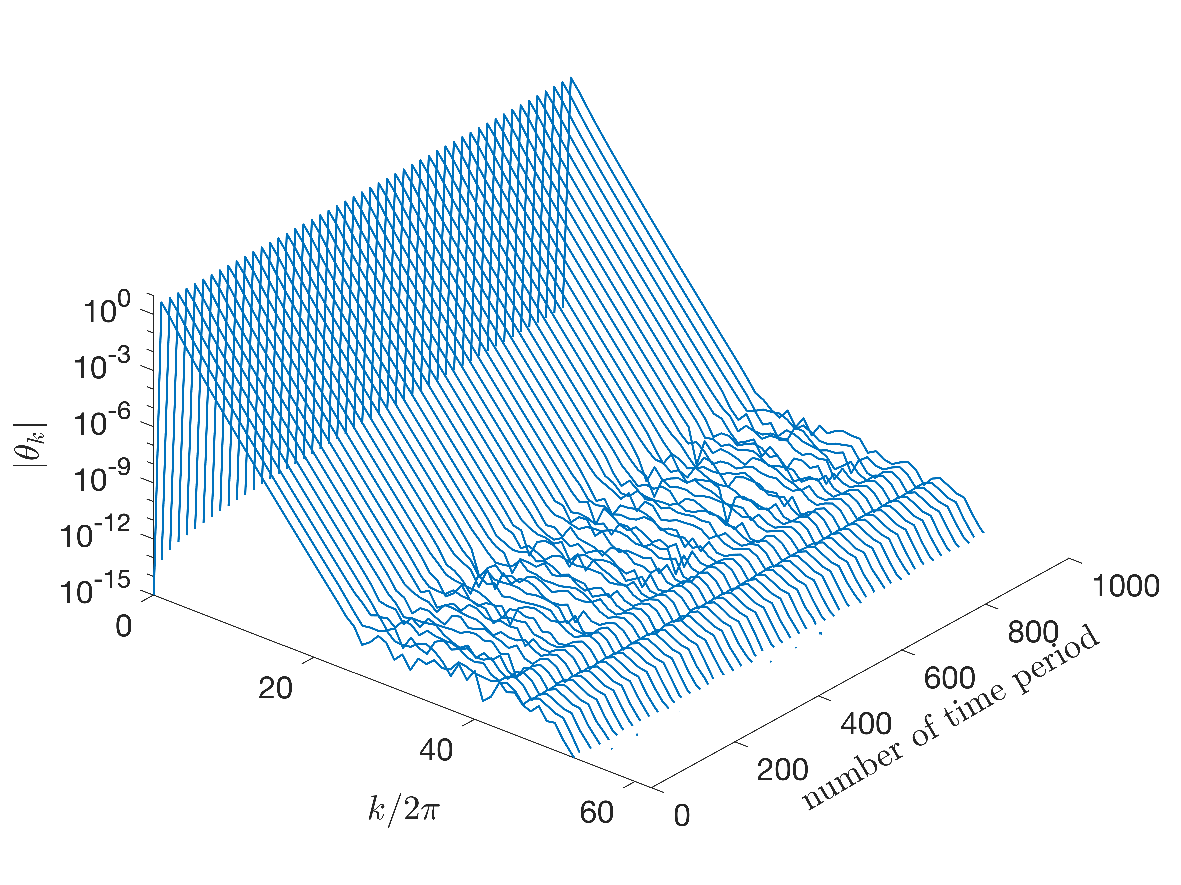}
    \end{minipage}
    }
    \caption{Propagation of a travelling surface gravity waves with $H=0.2$, $c\approx 0.912$ $k=1$, and $h_1 = 1$. (a) Top: wave profiles at $t = 0$ (blue) and $1000T$ (red), where $T$ is the temporal period. Middle: $err_z = |z(l,1000T)-z(l,0)|$, where $z = x(l,t)+\mathrm i\eta(l,t)$. Bottom: $E_r$ versus $t/T$. (b) $|\theta_k|$ versus $k/2\pi$ at different moments.}
    \label{fig:wave_propagation}
\end{figure}

Using the Newton-Krylov method described in section \ref{NewtonKyrlov}, we are able to obtain high-precision travelling wave solutions of surface gravity waves in finite water depth\footnote{We choose $h_1$, $\sqrt{gh_1}$, and $\sqrt{h_1/g}$, where $h_1$ denotes the water depth, as typical length, speed and time scales when dealing with pure gravity waves in finite water depth.}. We perform a long-term simulation to test our time-stepping algorithm using these solutions as initial conditions. Fig. \ref{fig:wave_propagation} shows the propagation of a gravity wave with crest-to-trough amplitude $H = 0.2$, wave speed $c\approx 0.912$, and wave number $k=1$. In this simulation, we set $N=128$ and $\Delta t = T/10000\approx 6.886\times 10^{-4}$, where $T$ is the temporal period. On the top of Fig. \ref{fig:wave_propagation}(a), we compare the wave profile at $t=1000T$ (red dots) with the profile at $t=0$ (blue curve). The difference between the two solutions is measured by
\begin{align}
    err_z := |z(l,1000T)-z(l,0)|,\label{err_z}
\end{align}
which is of order $10^{-11}$, as shown by the middle subfigure. The energy error $E_r$ is is plotted in bottom subfigure to show it is well controlled and less than $3\times 10^{-11}$ during the simulation. To suppress the aliasing error, we use the following $36$th-order Fourier filter\cite{hou2007computing}
\begin{align}
    f(k) = \mathrm e^{-36 (|k|/k_{max})^{36}}.\label{fourier filer}
\end{align}
In Fig. \ref{fig:wave_propagation}(b), we plot the spectrum of $\theta$. One can clearly see that there is no spurious growth in high-wave number region.
\subsubsection{Wave breaking}
\begin{figure}[!h]
    \centering
    \subfigure[]{
    \begin{minipage}{0.48\textwidth}
    \centering\includegraphics[width=\textwidth]{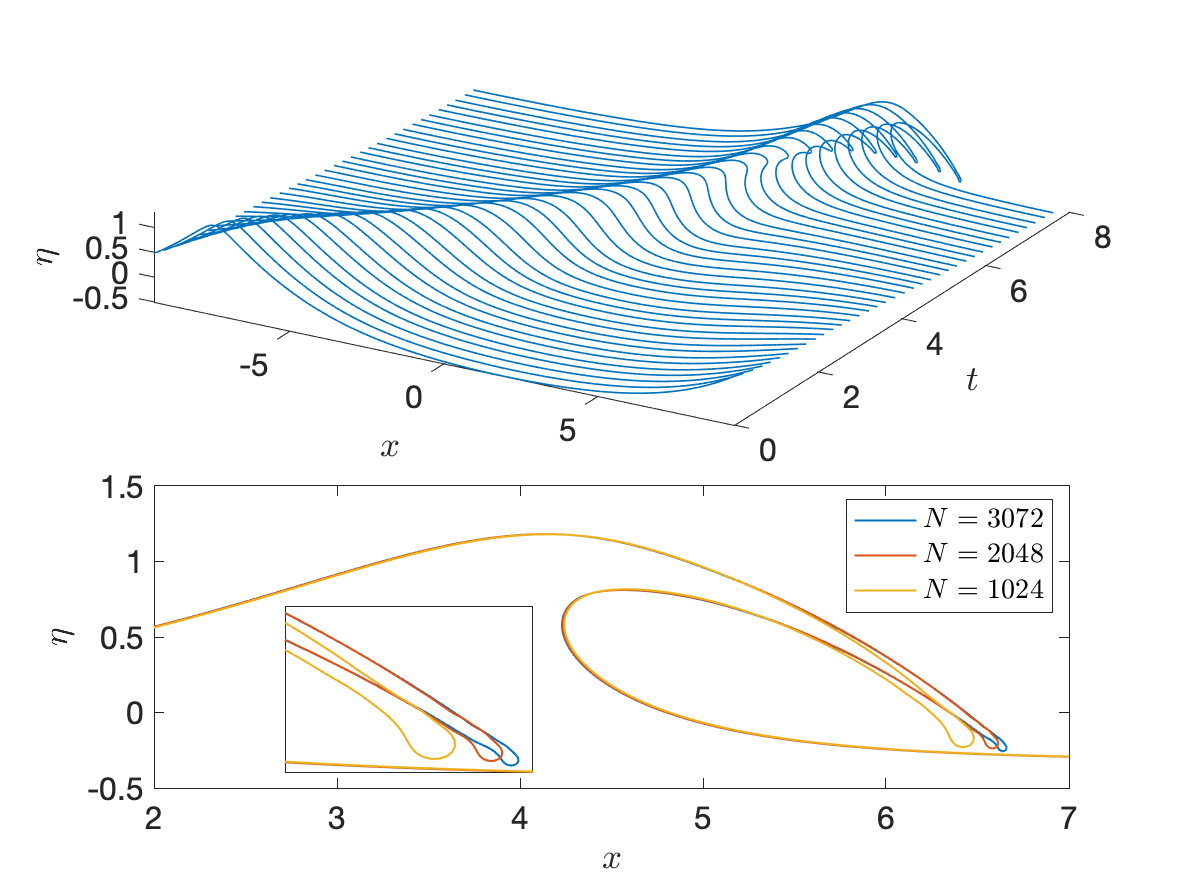}
    \end{minipage}
    }
    \subfigure[]{
    \begin{minipage}{0.48\textwidth}
    \centering\includegraphics[width=\textwidth]{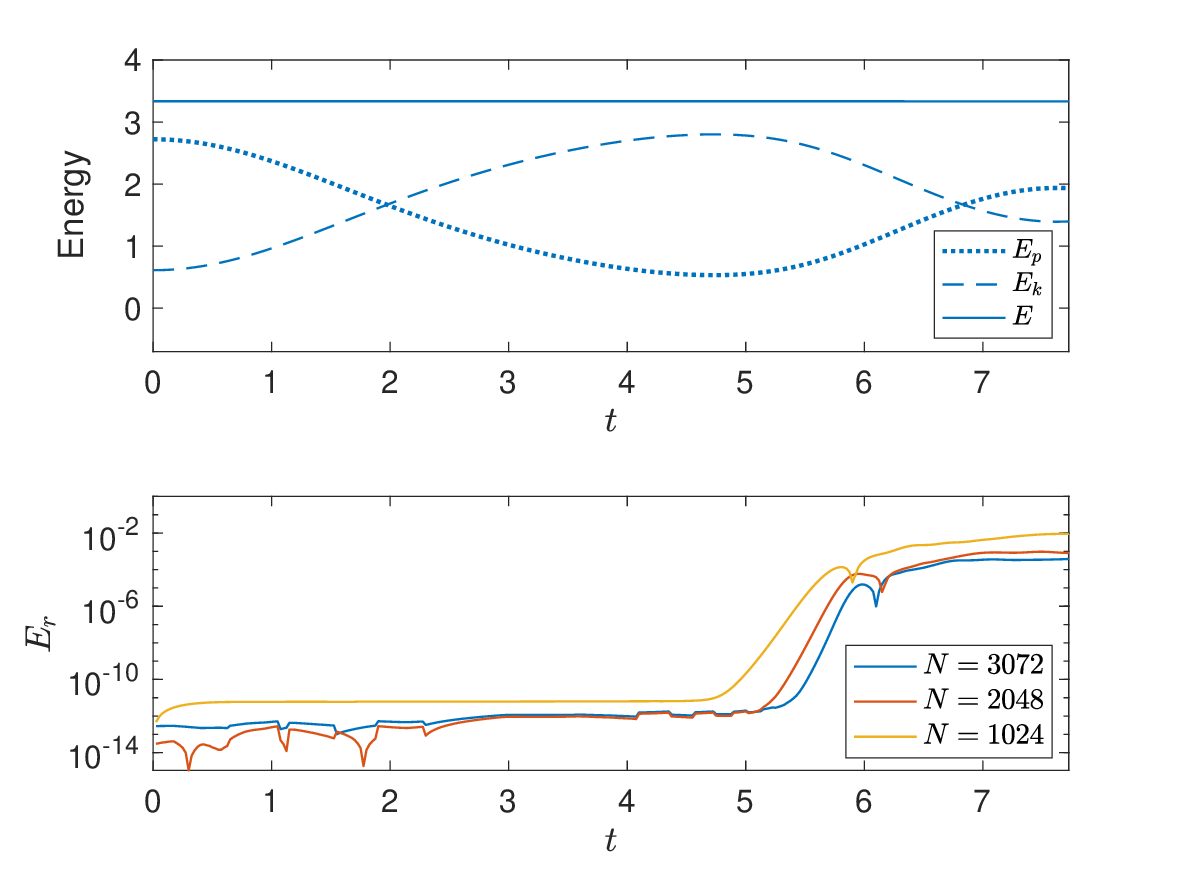}
    \end{minipage}
    }

    \subfigure[]{
    \begin{minipage}{0.48\textwidth}
    \centering\includegraphics[width=\textwidth]{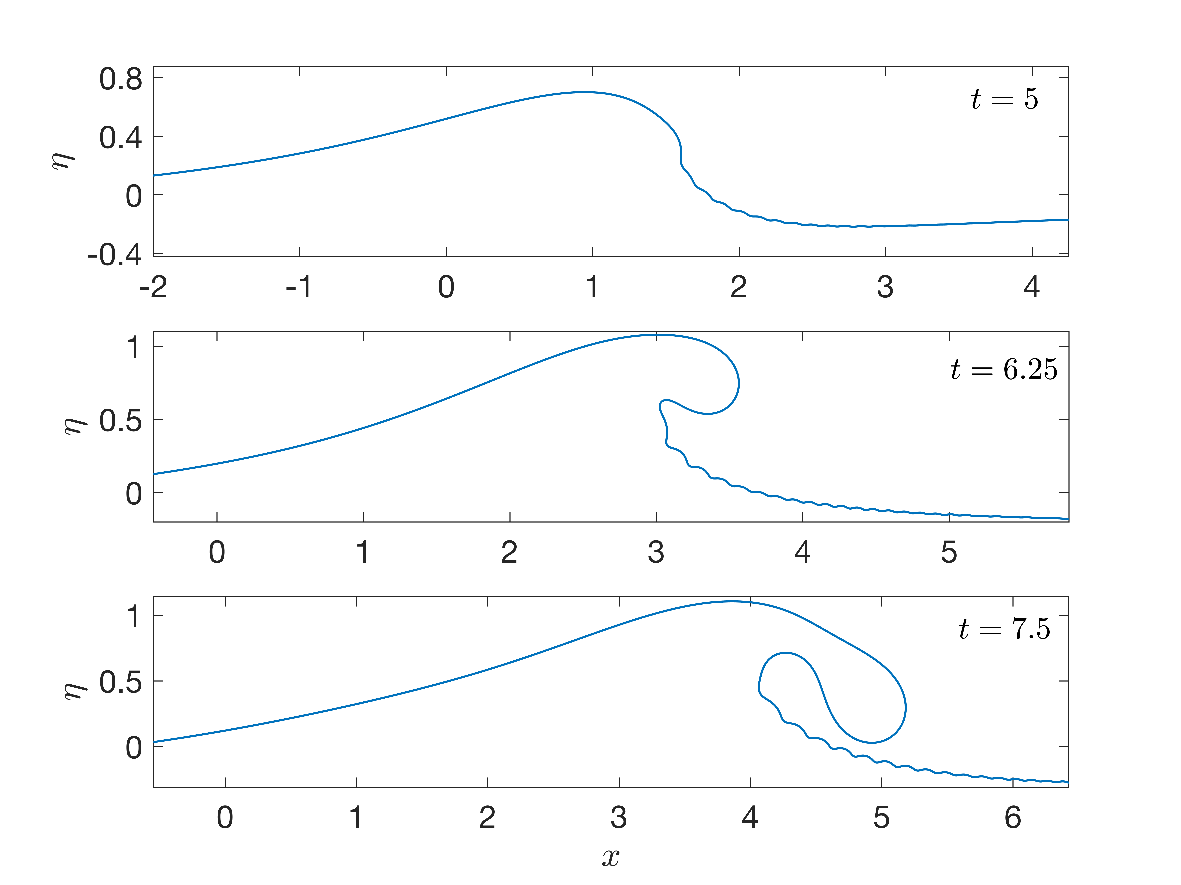}
    \end{minipage}
    }
    \subfigure[]{
    \begin{minipage}{0.48\textwidth}
    \centering\includegraphics[width=\textwidth]{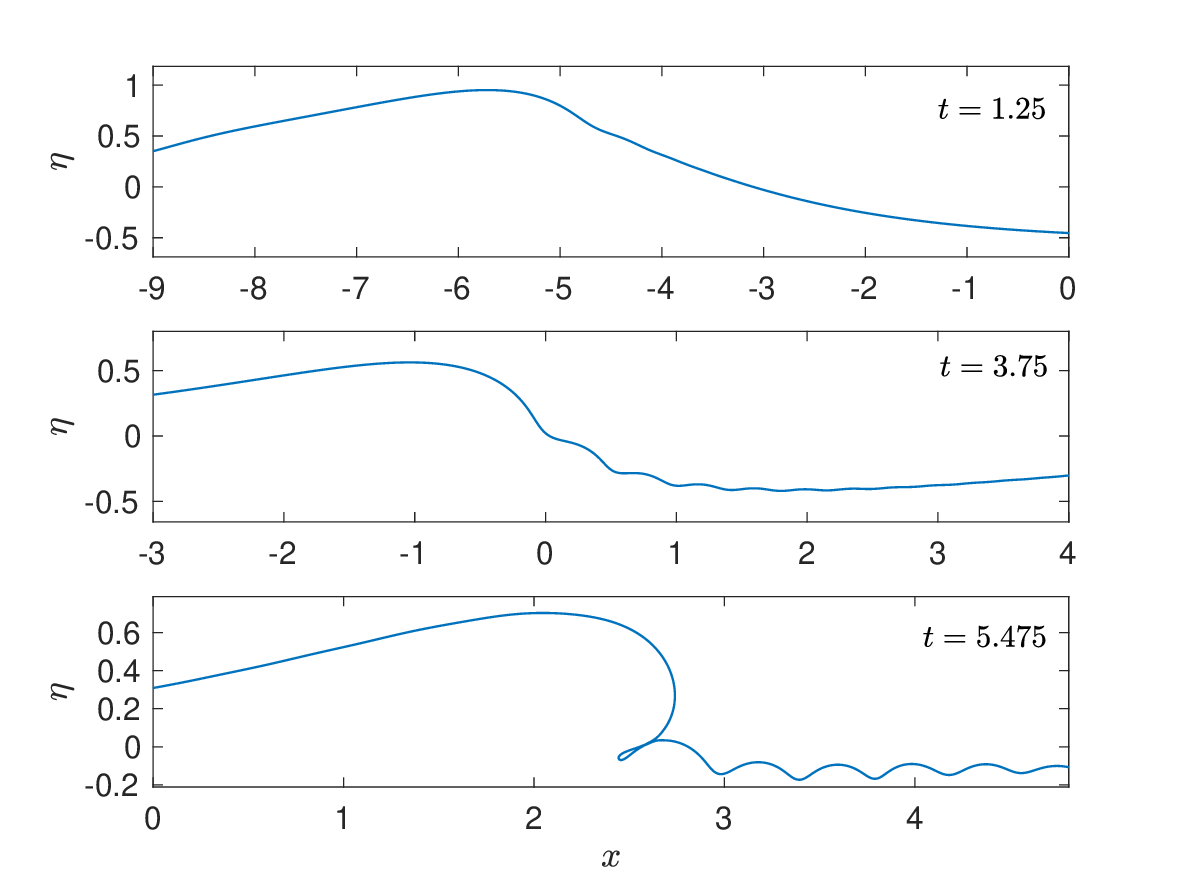}
    \end{minipage}
    }
    \caption{(a) Top: breaking process of a gravity wave. Bottom: wave profile at $t=7.725$ with $N=1024, 2048$ and $3072$. (b) Top: potential energy ($E_p$), kinetic energy ($E_k$) and total energy ($E$) versus $t$ for $N=3072$. Bottom: $E_r$ versus $t$ for $N=1024, 2048$ and $3072$. (c) Wave breaking process with surface tension ($Bo = 20$). (d) Air entrainment caused by surface tension ($Bo = 5$) before breaking happens.}
    \label{fig:break}
\end{figure}

To simulate wave breaking, we choose a travelling wave solution and amplify its  profile by a factor $\mu$. This is implemented by performing the following operations
\begin{align*}
    \theta(l)\rightarrow \theta(l),\quad \varphi(l)\rightarrow\varphi(l),\quad S\rightarrow \mu S, \quad x_0\rightarrow\mu x_0, \quad k\rightarrow k/\mu.
\end{align*}
Instead of using the Fourier filter, a $15$-point smoothing formula invented in \cite{dold1992efficient} turns out to be more robust when wave breaking happens. In Fig. \ref{fig:break}, a travelling wave solution having unit water depth with crest-to-trough amplitude $H = 0.6$ and wave number $k=1$ is amplified by factor $\mu = 3$ and set to be the initial condition. Time step $\Delta t = 5\times 10^{-4}$.
At the initial stage, gravitational potential energy $E_p$ is transferred to kinematic energy $E_k$, resulting an accelerating moving front. At $t\approx 5.375$, the wave develops a vertical tangent and then becomes overhanging. A plunging breaker appears and becomes sharp at its tip. At $t=7.725$, the wave becomes almost self-touching and encloses an air bubble on its front face. The nipple of the breaker has a locally round shape and tiny wiggles, which probably come from numerical oscillations due to the drastic increase of curvature. Higher resolutions are used to show that a gradually convergent solution can be obtained. A comparison of the wave profile at $t=7.725$ with $N=1024, 2048$ and $3072$ is plotted on the bottom of Fig. \ref{fig:break}(a). The total energy is shown to be a constant between $3.334$ and $3.335$. At $t\approx 4.75$, the kinetic energy $E_k$ and gravitational potential energy $E_p$ reach their extremums. Shortly after that, wave becomes overturned and the relative energy error $E_r$ increases rapidly, as shown in Fig. \ref{fig:break}(b). When $N = 1024, 2048$ and $3072$, the upper bounds of $E_r$ are and $10^{-2}, 10^{-3}$ and $3\times 10^{-4}$ respectively, showing a convergent behaviour of solution.
On the other hand, it is found that the wave breaking process can be significantly influenced by capillary effect, which generates small-scale ripples, known as parasitic waves. Surface tension effect, measured by the Bond number $Bo$, can modify the characteristics of breakers, stabilize, and even suppress the breaking process. In Fig. \ref{fig:break}(c) and (d), we perform simulations using the same initial condition as the previous one with the Bond number $Bo = 20$ and $5$. They represent relatively weak and strong surface tension effects. When $Bo = 20$, a large-scale plunging breaker is ultimately generated. However, the jet becomes larger and the tip is more round compared with gravitational plunging breaker. When $Bo = 5$, capillary effect is strong enough to suppress the formation of jet although a small overhanging structure is observed. In the end of simulation, we observe a closed air bubble, having a profile similar to the Crapper waves. This is typical for spilling breakers which characterize air entrainment and bubbles \cite{deike2015capillary}. 

\subsubsection{Head-on collision of solitary waves}
\begin{figure}[!h]
    \centering
    \subfigure[]{
    \begin{minipage}{0.48\textwidth}
    \centering\includegraphics[width=\textwidth]{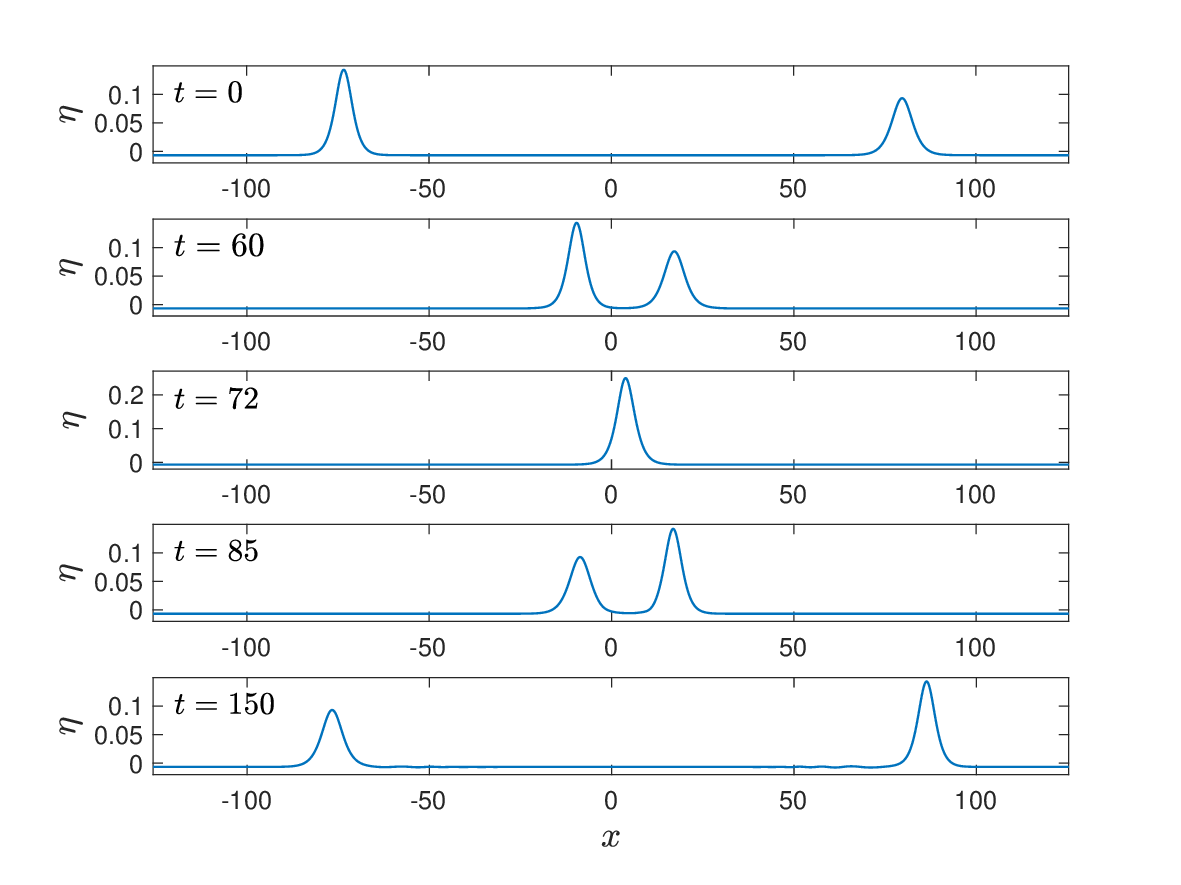}
    \end{minipage}
    }
    \subfigure[]{
    \begin{minipage}{0.48\textwidth}
    \centering\includegraphics[width=\textwidth]{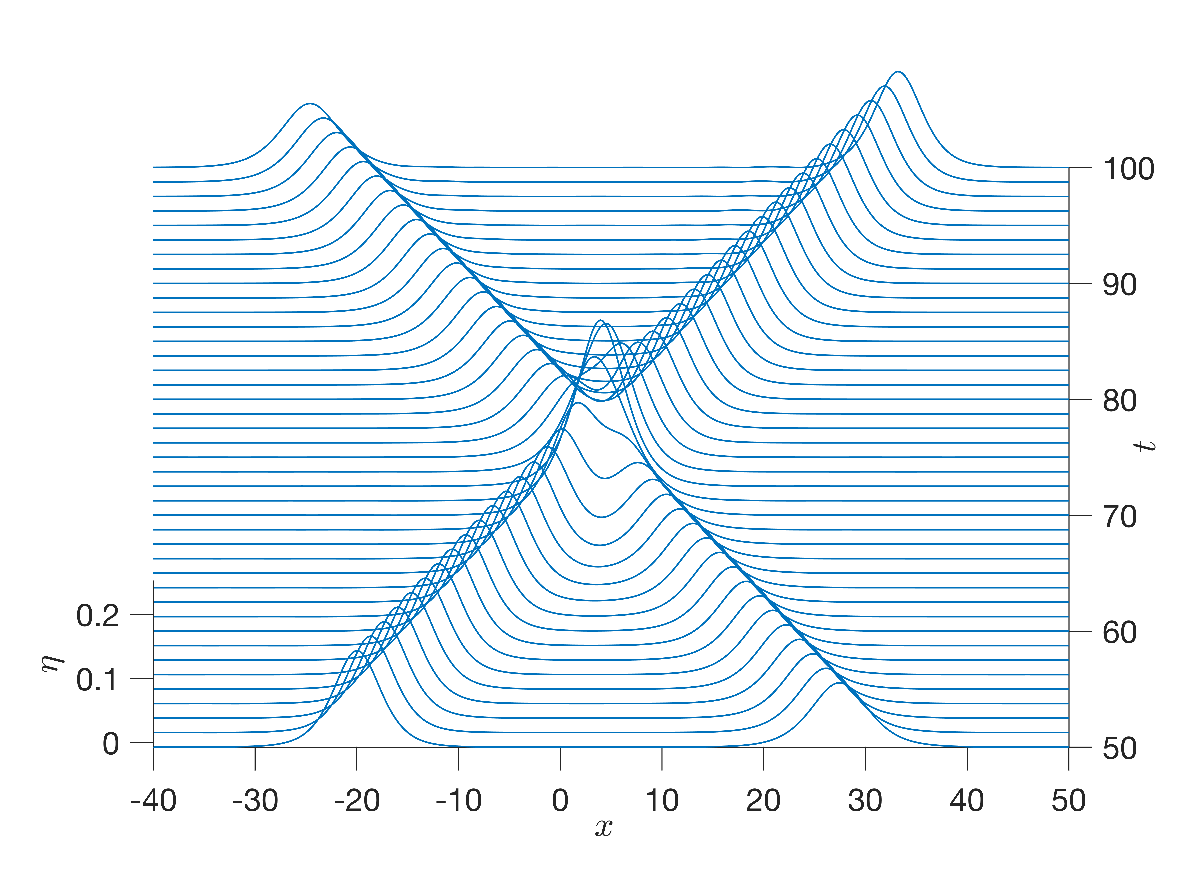}
    \end{minipage}
    }
    \caption{Head-on collision of two solitary waves with wave amplitude $0.1$ and $0.15$. (a) Profiles at $t = 0, 60, 72, 85$, and $150$. (b) Local profiles for $50<t<100$.}
    \label{fig:solitary collide}
\end{figure}

In Fig. \ref{fig:solitary collide}, we exhibit a simulation of head-on collision of two solitary waves. Setting $k = 0.025$, we can simulate solitary waves by using long periodic waves. The initial condition is a superposition of two well-separated solitary waves with amplitude $H = 0.1, 0.15$ and opposite travelling direction. In the simulation, we choose $N=1024$, $\Delta t = 10^{-3}$, and apply the $36$th-order Fourier filter. The profiles at $t=0, 60, 72, 85$, and $150$ are shown in Fig. \ref{fig:solitary collide}(a). As expected, the two solitary waves maintain their shape after collision, and radiate some tiny wave trains on their rear faces. A local illustration of the collision process is shown in Fig. \ref{fig:solitary collide}(b). During the simulation, the energy error $E_r$ is well controlled and less than $2\times 10^{-11}$.

\subsection{Interfacial gravity-capillary solitary waves}
\subsubsection{Bifurcations and profiles}
It is well-known that the envelop of small-amplitude interfacial gravity-capillary solitary waves $A(x,t)$ satisfies the NLS equation \cite{laget1997numerical}
\begin{align}
    \mathrm i A_t + \alpha A_{xx} + \beta |A|^2A = 0,
\end{align}
with
\begin{align}
    \alpha = \frac{\omega}{4(1-R)},\quad \beta = \frac{\omega (1-R)}{2}\bigg[ \frac{4(1-R)^2}{(1+R)^2}-\frac{5}{4} \bigg],
\end{align}
where $\omega$ is defined in (\ref{c_p}).
There exists a critical density ratio $R_c=(21-8\sqrt 5)/11\approx 0.283$. When $R<R_c$, the NLS equation is of the focussing type and has localised solitary wave solutions bifurcating from infinitesimal periodic waves. When $R>R_c$, the NLS equation is of the defocusing type and only has non-localised dark soliton solutions. However, solitary waves of Euler equations have been found even when the corresponding NLS equation is of the defocusing type \cite{laget1997numerical}. These solitary waves can only exist with finite amplitude, thus have no contradiction to the weakly nonlinear theory. Similar situation is also encountered in hydroelastic waves which model the motion of ice sheet \cite{milewski2011hydroelastic}. Recent studies have shown that when $R>R_c$, solitary waves evolve from the dark solitons or the generalised solitary waves, which themselves bifurcate from periodic waves \cite{milewski2011hydroelastic,wang2014quasi}. In this section, we focus on the stability of solitary waves and leave the dark solitons in future studies.

\begin{figure}[!h]
    \centering
    \subfigure[]{
    \begin{minipage}{0.48\textwidth}
    \centering\includegraphics[width=\textwidth]{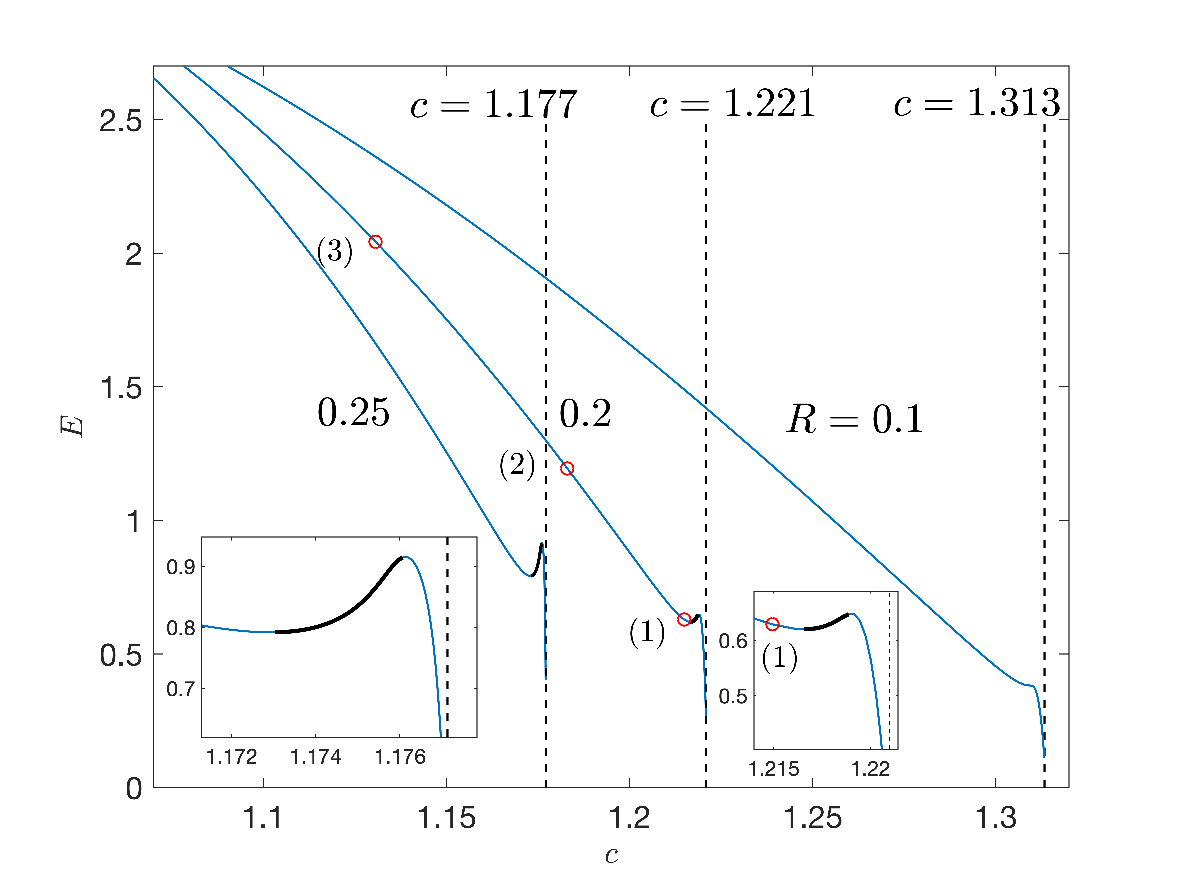}
    \end{minipage}
    }
    \subfigure[]{
    \begin{minipage}{0.48\textwidth}
    \centering\includegraphics[width=\textwidth]{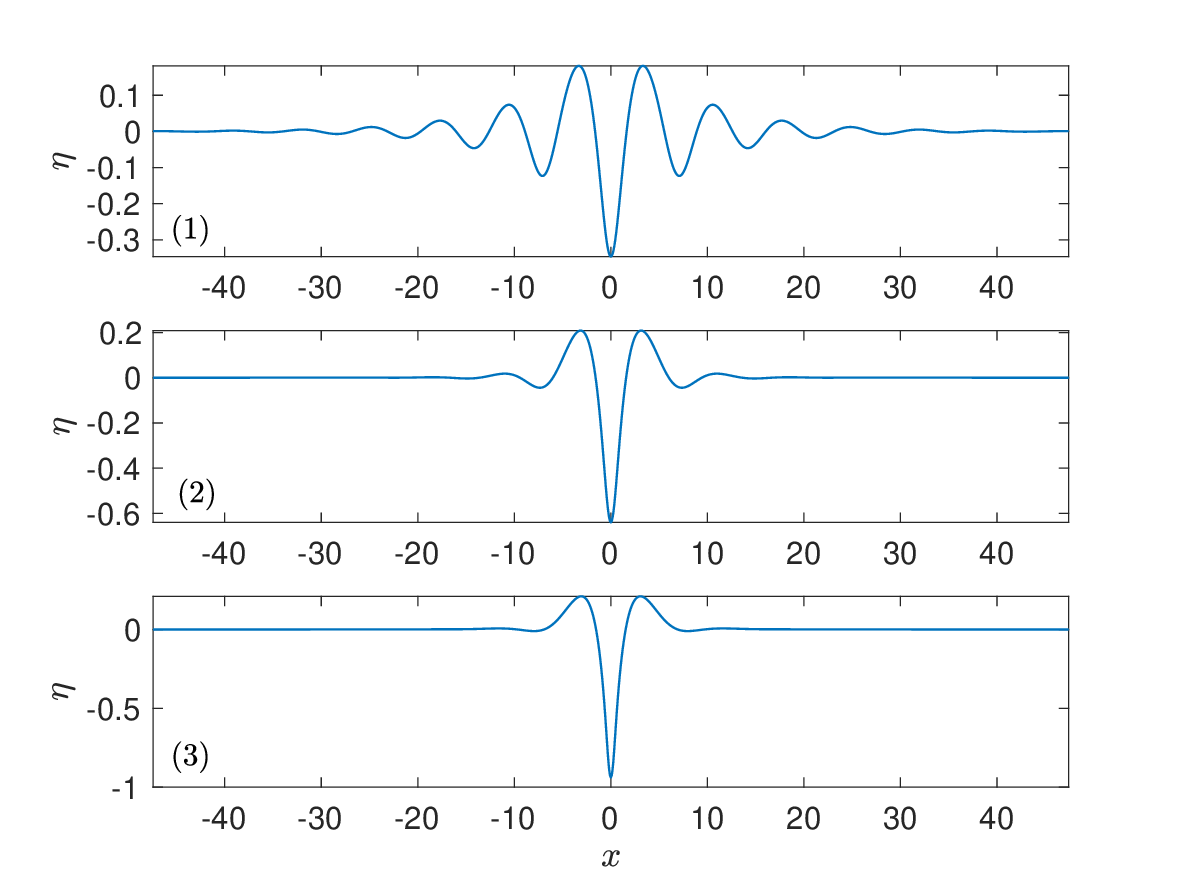}
    \end{minipage}
    }
    \caption{(a) Speed-energy bifurcation curves of depression solitary waves for $R=0.1, 0.2$ and $0.25$. The black dashed lines label the minimum phase speed where solitary waves bifurcate. In the small windows, a local blow-up shows the non-monotonic behaviour of the speed-energy curve. The black curve on the bifurcations represents linearly unstable solutions. (b) Three typical waves, labelled by the red circles, along the branch of $R=0.2$.}
    \label{fig:bif1}
\end{figure}

In Fig. \ref{fig:bif1} (a), we plot three speed-energy bifurcations of depression solitary waves with $R=0.1, 0.2$ and $0.25$. They belong to the case $R<R_c$. Therefore, solitary waves bifurcate from infinitesimal periodic waves at the minimum phase speed $c_{min} = \sqrt{2}(1-R)^{1/4}/(1+R)^{1/2}$. At small amplitude, they have wave-packet profiles with spatially decaying tails in the far field. When wave amplitude increases, the wave speed decreases monotonically. We display three typical waves with $R=0.2$ in Fig. \ref{fig:bif1} (b). Ultimately, only one single trough at the center survives and keeps steepening until an overhanging structure develops. The limiting configuration of the depression solitary waves is a self-intersecting wave with a closed fluid bubble at the trough, similar to the famous Crapper waves locally \cite{crapper1957exact}.

\begin{figure}[!h]
    \centering
    \subfigure[]{
    \begin{minipage}{0.48\textwidth}
    \centering\includegraphics[width=\textwidth]{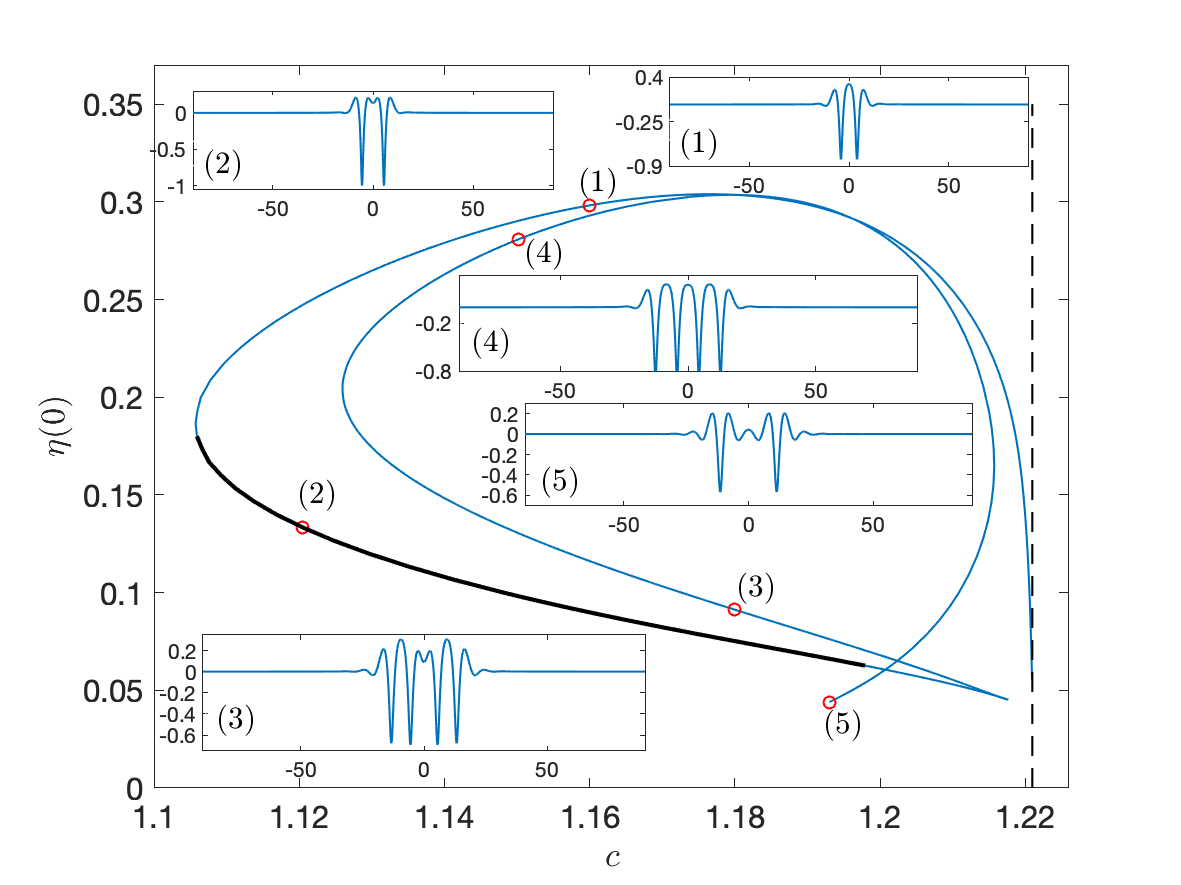}
    \end{minipage}
    }
    \subfigure[]{
    \begin{minipage}{0.48\textwidth}
    \centering\includegraphics[width=\textwidth]{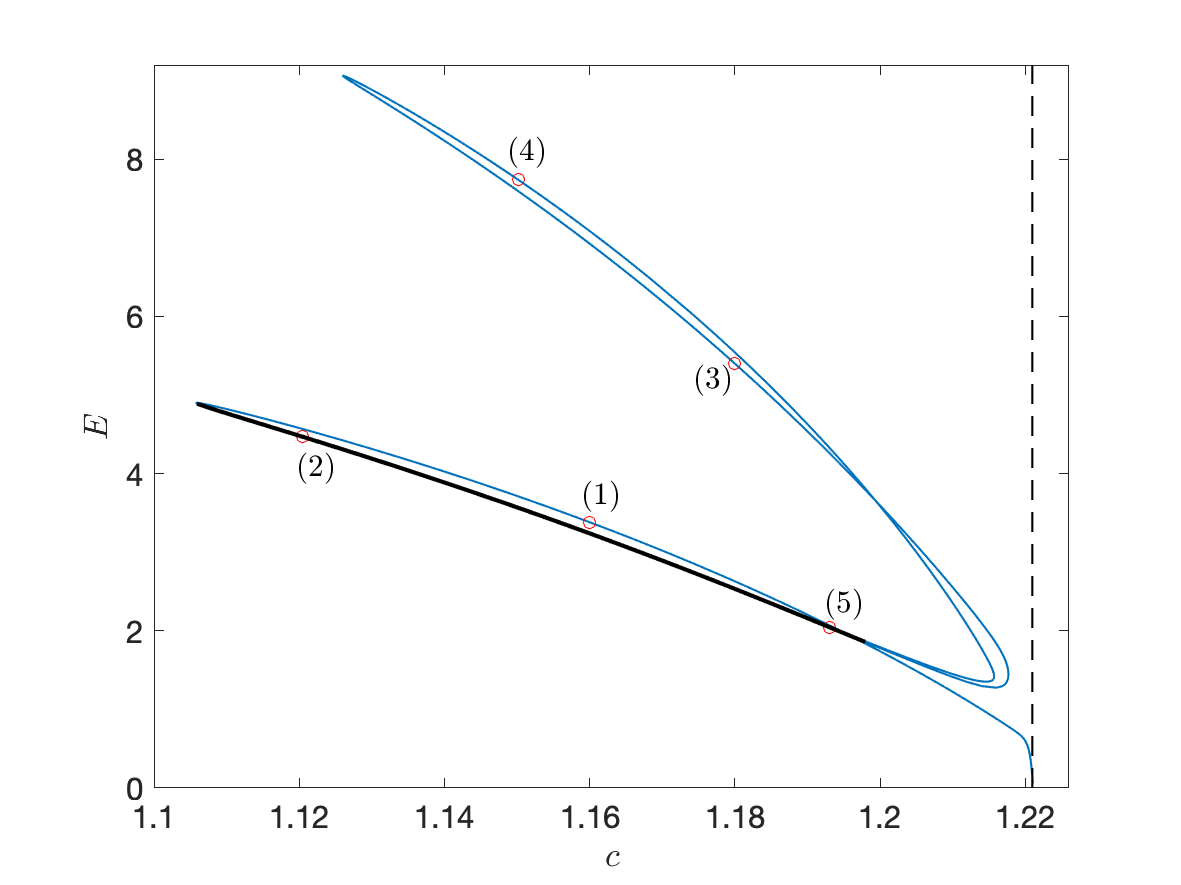}
    \end{minipage}
    }
    \caption{(a) Speed-amplitude bifurcation curve and five typical solutions of elevation solitary waves for $R=0.2$. (b) Speed-energy bifurcation of elevation solitary waves for $R=0.2$. The black dashed lines label the minimum of phase speed. The black solid curves represent linearly stable solutions.}
    \label{fig:ele_bif}
\end{figure}

On the other hand, the evolution of elevation solitary waves is more complicated.  In Fig. \ref{fig:ele_bif} (a), we plot the speed-amplitude bifurcation curve of $R=0.2$, as well as five typical waves. In Fig. \ref{fig:ele_bif} (b), we plot the corresponding speed-energy curve. Close to the bifurcation point $c\approx 1.221$, elevation solitary waves feature wave-packet profiles with positive values of $\eta$ at the center. When the wave amplitude increases, wave speed decreases monotonically until the first turning point appears. Depression solitary waves then gradually develop more troughs and peaks following the bifurcation . After passing through four turning points, they resemble two side-by-side depression waves. Readers who are interested in the global bifurcation structure are referred to \cite{doak2022global} for details.

\subsubsection{Linear stability}
Using the new formulation described in section 4.3.2, we perform linear stability analysis to solitary waves with $R=0.2$. As shown in Fig. \ref{fig:bif1}, the energy curve of the depression solitary waves undergoes two stationary points. According to \cite{saffman1985superharmonic}, this implies that exchange of stability could happen at these points. Indeed, we found that depression solitary waves are linearly stable except the those solutions between the two stationary points. The linearly unstable solutions are represented by the black solid lines in Fig. \ref{fig:bif1} (a). This is different from the case $R = 0$, i.e. depression gravity-capillary solitary waves of surface type, which have a monotonic energy curve and are linearly stable no matter their amplitude. 

\begin{figure}[!h]
    \centering
    \subfigure[]{
    \begin{minipage}{0.48\textwidth}
    \centering\includegraphics[width=\textwidth]{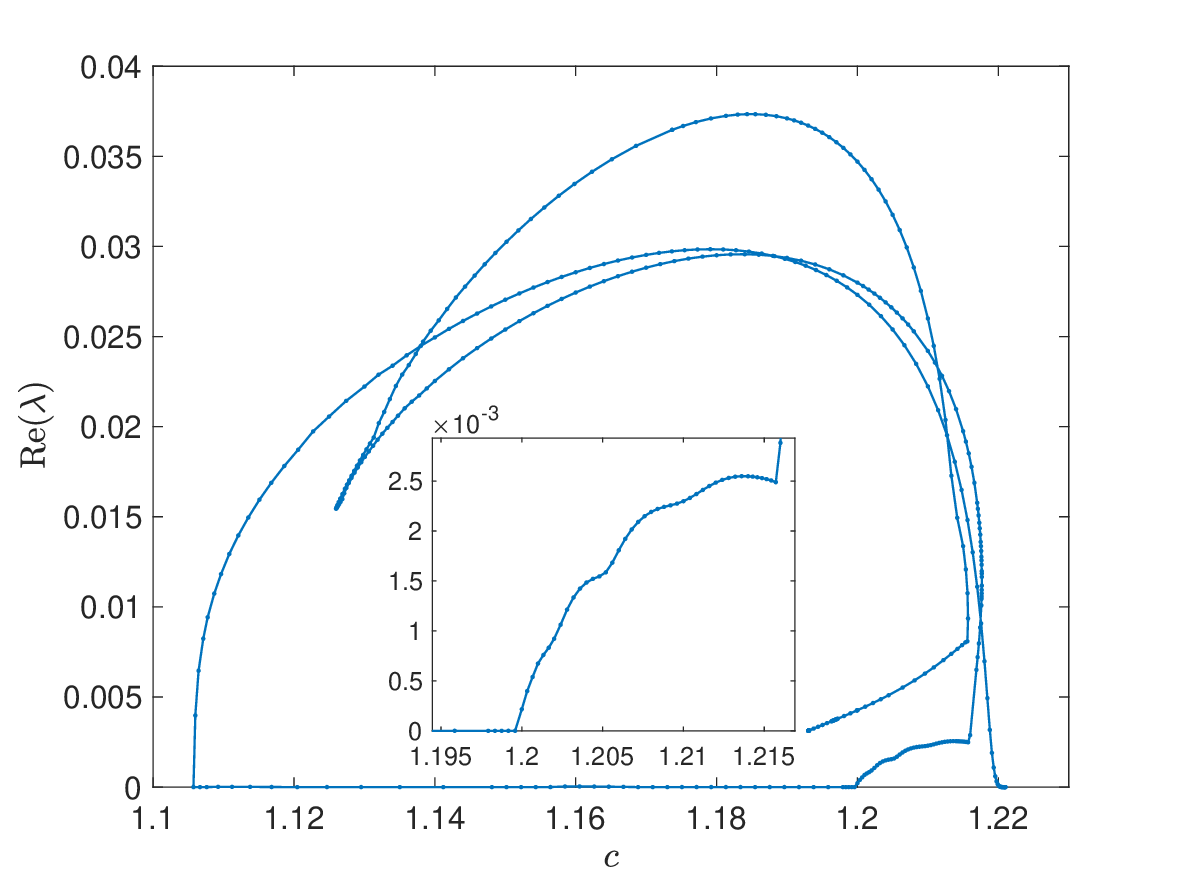}
    \end{minipage}
    }
    \subfigure[]{
    \begin{minipage}{0.48\textwidth}
    \centering\includegraphics[width=\textwidth]{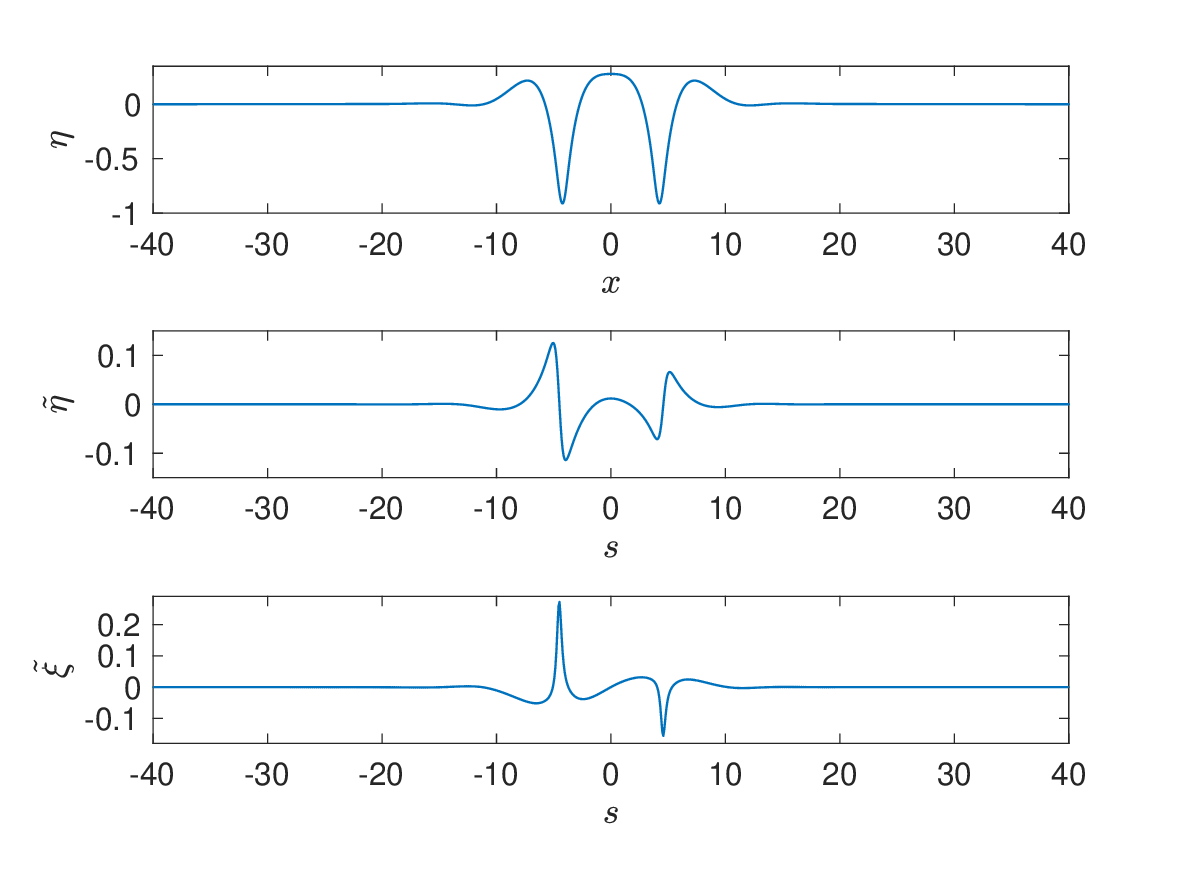}
    \end{minipage}
    }
    \caption{(a) $\text{Re}(\lambda)$ versus $c$ for the elevation branch of solitary waves of $R=0.2$. (b) Top: elevation solitary wave of $R=0.2$ with $c = 1.14$ and $\eta(0) \approx 0.279$. Middle: $\tilde \eta$ corresponding to the most unstable mode. Bottom: $\tilde \xi$ corresponding to the most unstable mode.}
    \label{fig:ele_unstable}
\end{figure}

For elevation solitary waves, on the other hand, small-amplitude solutions are linearly unstable. As there exist multiple stationary points on the energy curve, the first exchange of stability is found at $c\approx 1.106$. Solutions then becomes linearly stable until $c\approx 1.2$, as shown in Fig. \ref{fig:ele_bif}. Surprisingly, this is not a stationary point of energy. In fact, what happens there is a dominant superharmonic instability instead of the excahnge of stability. The difference between the two concepts is that the former has nonzero imaginary part of growth rate, i.e. $\text{Im}(\lambda)\neq 0$, while the latter has zero imaginary part of growth rate. The linear growth rate $\text{Re}(\lambda)$ is plotted in Fig.\ref{fig:ele_unstable}(a).  At $c\approx 1.216$, i.e. the second stationary point of energy, an exchange of stability happens and replaces the superharmonic instability, which is manifested by the sharp angle of the curve. When $R=0$, i.e. surface gravity-capillary solitary waves, we also observe that the superharmonic instability emerges and dominates before the second stationary point appears.

\begin{figure}[!h]
    \centering
    \subfigure[]{
    \begin{minipage}{0.48\textwidth}
    \centering\includegraphics[width=\textwidth]{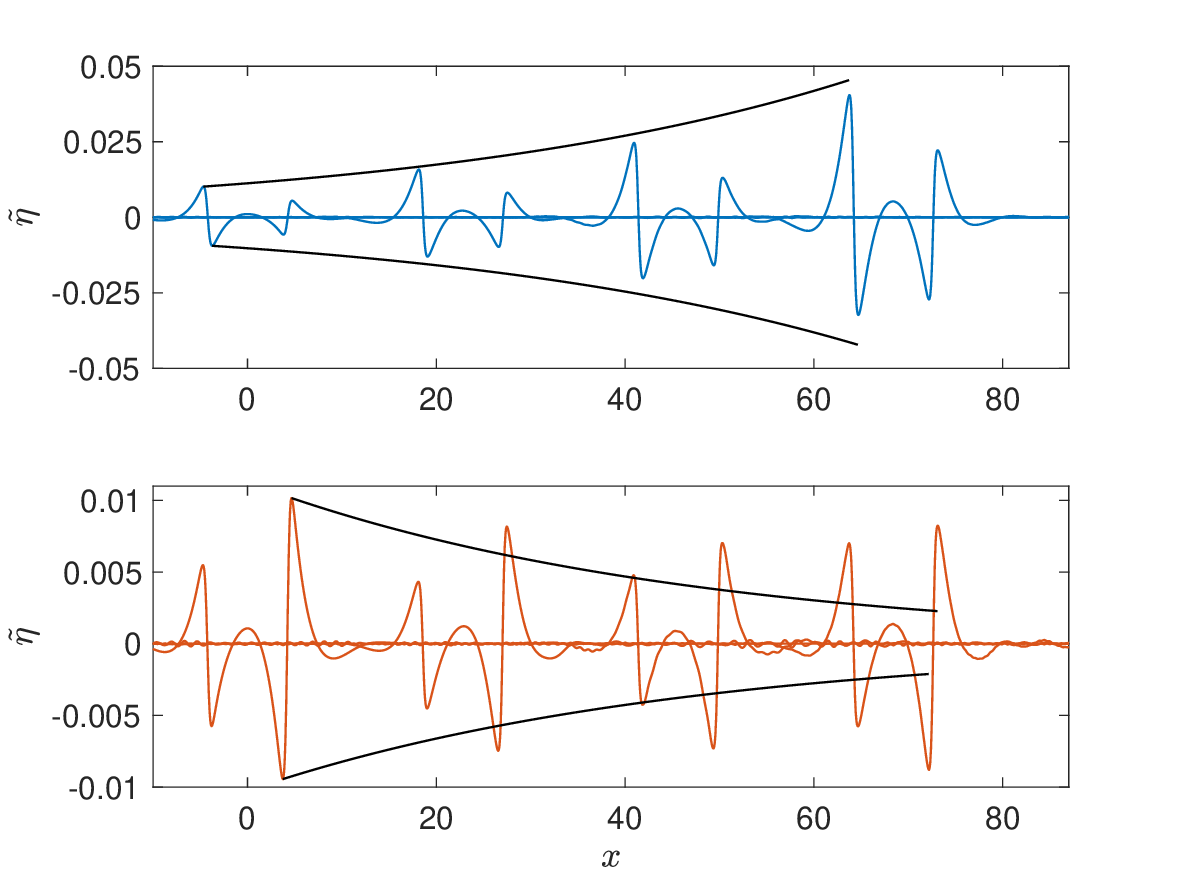}
    \end{minipage}
    }
    \subfigure[]{
    \begin{minipage}{0.48\textwidth}
    \centering\includegraphics[width=\textwidth]{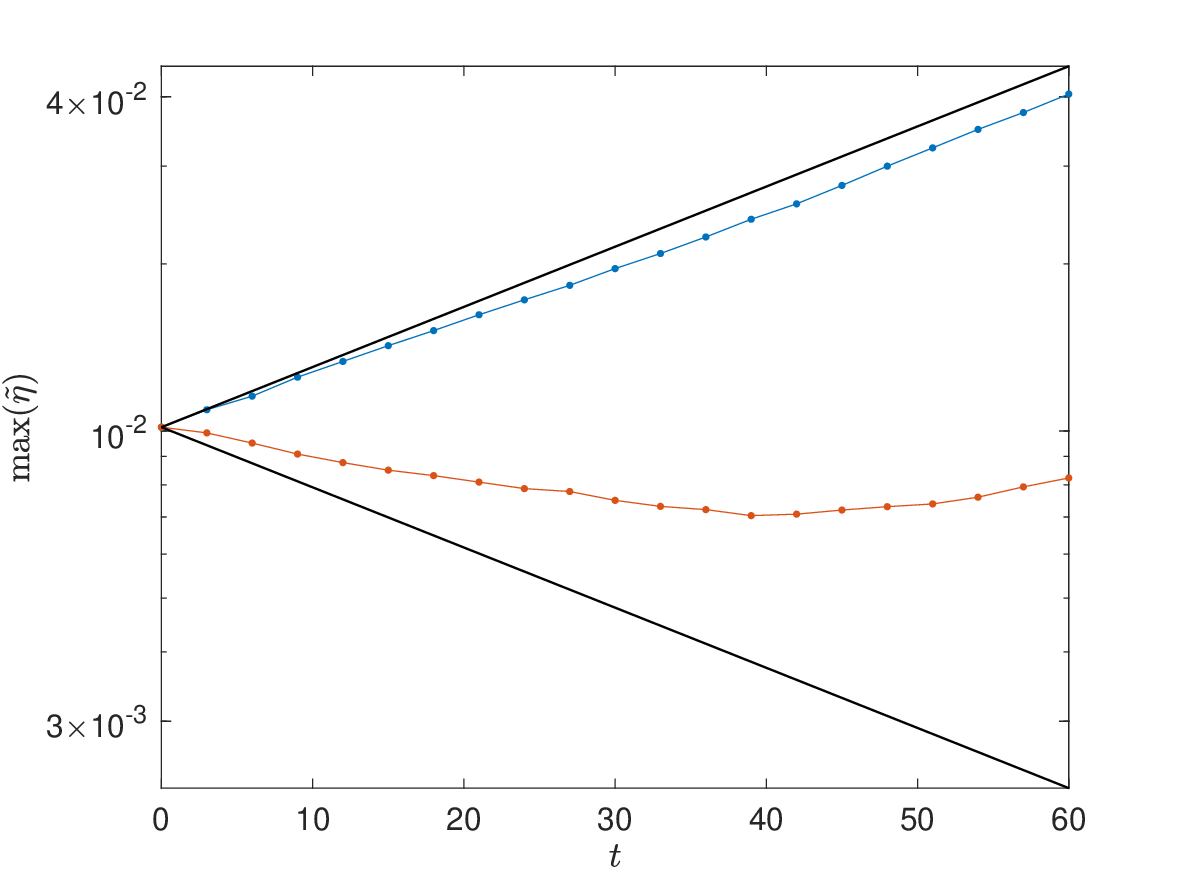}
    \end{minipage}
    }
    \caption{(a) Evolution of $\tilde\eta$ at $t=0, 20, 40$, and $60$ from left to right. Disturbances at $t=0$ correspond to the mode with growth rate $\lambda\approx0.02496$ (top) and $-0.02496$ (bottom). (b) max($\tilde\eta$) as a function of $t$ corresponding to the growing (blue) and decaying (red) mode, as well as the prediction based on linear stability analysis (black).}
    \label{fig:linear compare1}
\end{figure}

In Fig. \ref{fig:ele_unstable}(b), a linearly unstable elevation solitary wave with wave speed $c = 1.14$ and wave height $\eta(0) \approx 0.279$ is plotted, as well as its most unstable modes for $\tilde\eta$ and $\tilde\xi$. By setting $N = 2048$, and $4096$, we get the growth rate $\lambda = 0.024951$, and $0.024956$ respectively. 
To verify the linear stability theory, we perform fully nonlinear simulation for this solution. The solitary wave is superposed a small perturbation which is proportional to the most unstable mode and then set to be the initial condition. The evolution of the disturbance $\tilde\eta$ is plotted on the top of Fig. \ref{fig:linear compare1} (a). The profiles from left to right correspond to $t=0, 20, 40$, and $60$. The prediction from linear stability analysis is shown by the black curve, which represents the envelope of $\tilde\eta$ and exhibits good agreement to the fully nonlinear simulation. It is interesting to note that there exists a linearly stable mode whose growth rate equals to $-\lambda$. On the bottom of Fig. \ref{fig:linear compare1} (a), we show the time evolution of this mode when it is superposed to the solitary wave as perturbatoin. Profiles of $\tilde\eta$ are plotted at $t=0, 20, 40$, and $60$ from left to right. Initially $\tilde\eta$ decreases roughly following the prediction of linear stability until after $t=40$ the disturbance starts to grow. The growth rate of the two modes based on fully nonlinear calculation and linear stability theory are shown in Fig. \ref{fig:linear compare1} (b). Initially, linear stability theory predicts the growth rate of the most unstable mode quite well, but slightly overestimates it later. For the linearly stable mode, however, linear theory does not give a very good estimation. 

\begin{figure}[!h]
    \centering
    \subfigure[]{
    \begin{minipage}{0.48\textwidth}
    \centering\includegraphics[width=\textwidth]{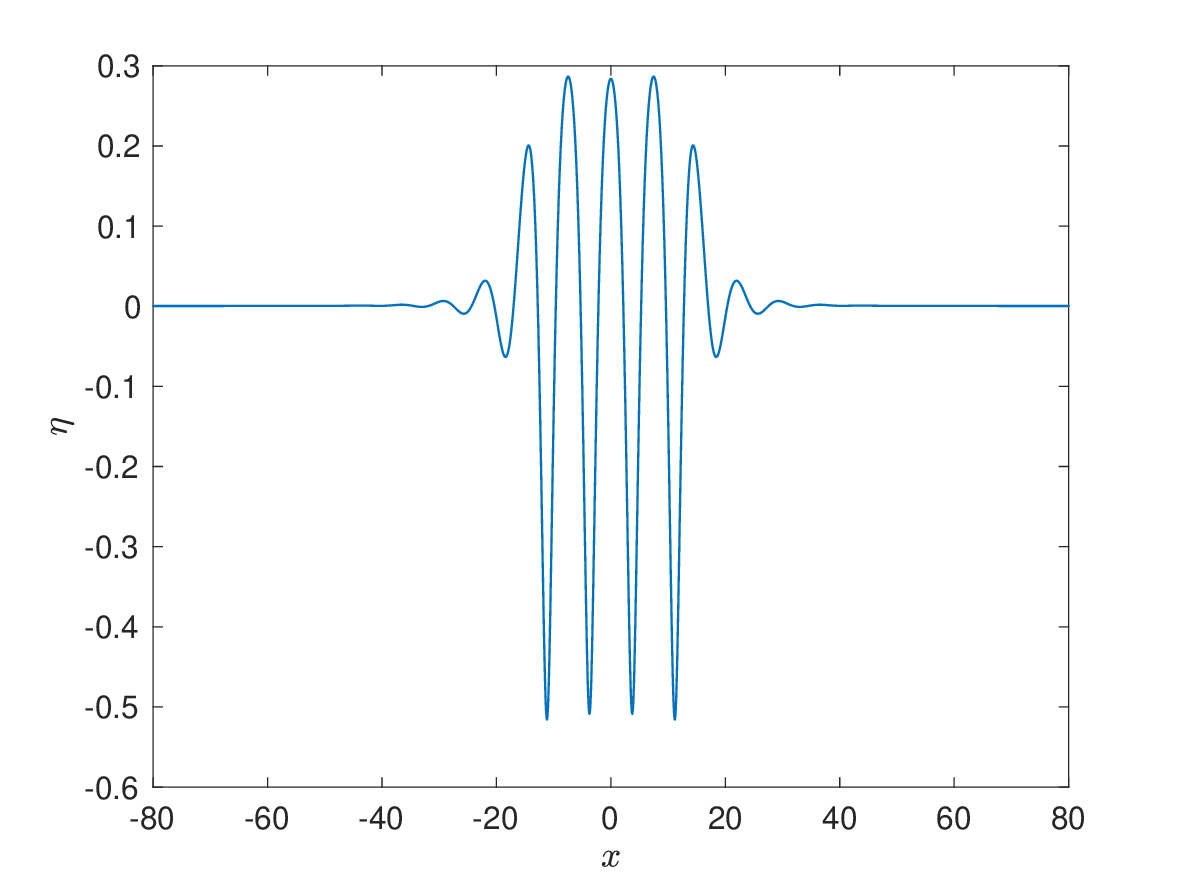}
    \end{minipage}
    }
    \subfigure[]{
    \begin{minipage}{0.48\textwidth}
    \centering\includegraphics[width=\textwidth]{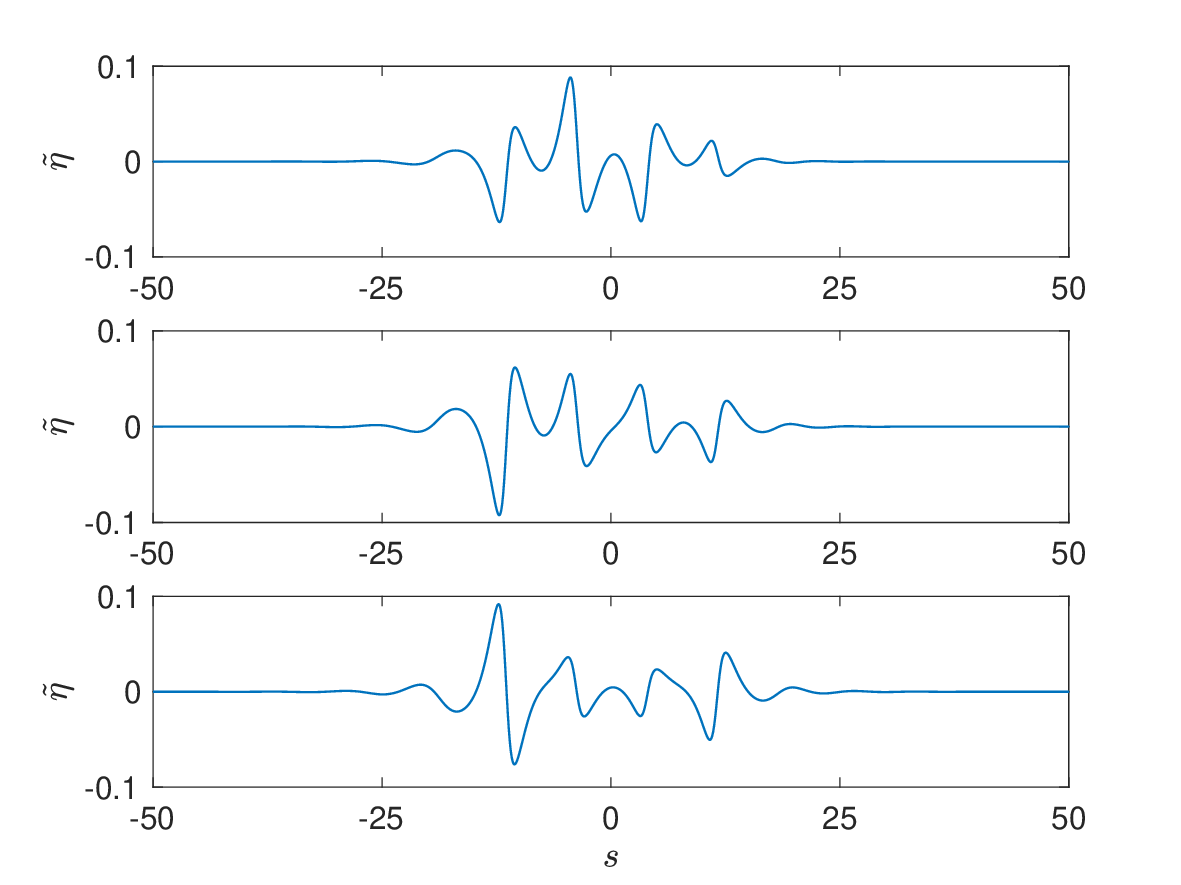}
    \end{minipage}
    }
    \caption{(a) Elevation solitary waves of $R=0.2$ with $c=1.2$ and $\eta(0) \approx 0.2836$. (b) Three unstable modes of $\tilde\eta$ with grow rates $\lambda \approx 0.034705, 0.026087, 0.013585$ from top to bottom. }
    \label{fig:3 modes}
\end{figure}

\begin{figure}[!h]
    \centering
    \subfigure[]{
    \begin{minipage}{0.48\textwidth}
    \centering\includegraphics[width=\textwidth]{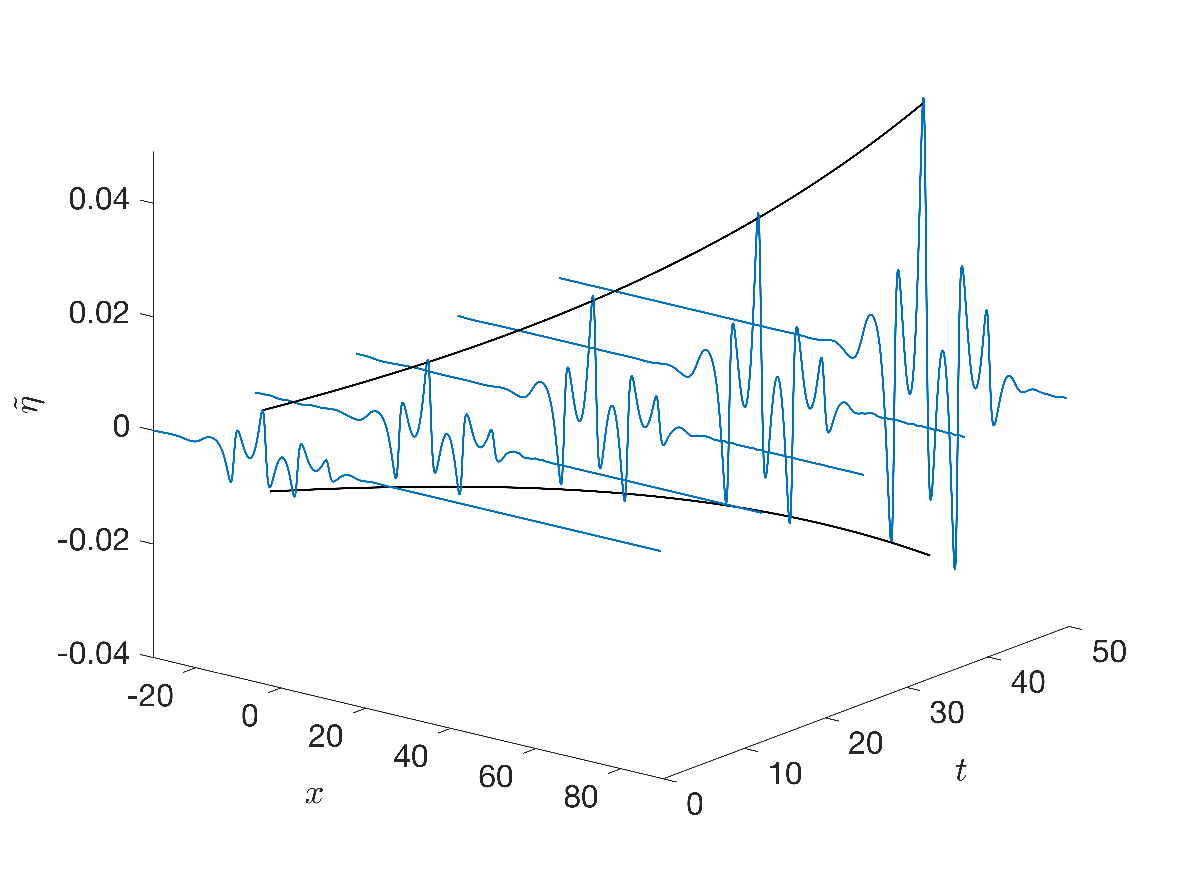}
    \end{minipage}
    }
    \subfigure[]{
    \begin{minipage}{0.48\textwidth}
    \centering\includegraphics[width=\textwidth]{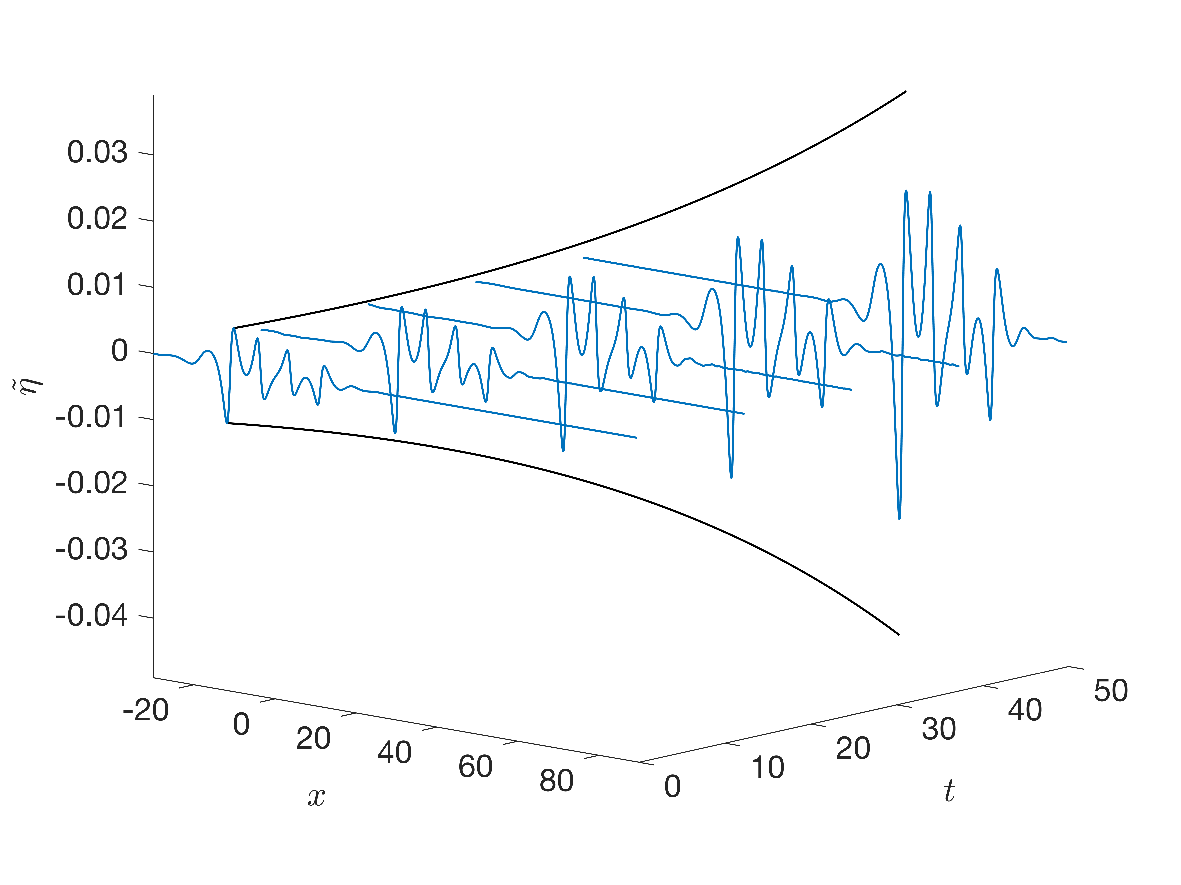}
    \end{minipage}
    }
    \caption{(a) Evolution of $\tilde\eta$ for the most unstable mode with $\lambda \approx 0.034705$. (b) Evolution of $\tilde\eta$ for the second most unstable mode with $\lambda \approx 0.026087$. The blue profiles and black envelopes are from nonlinear simulation and linear stability analysis.}
    \label{fig:linear compare2}
\end{figure}

When the exchange of stability happens, solutions can have more than one unstable modes. In Fig. \ref{fig:3 modes} (a), we display a solitary wave with $R = 0.2, c = 1.2$ and $\eta(0) \approx 0.2836$. It is located on the linearly unstable branch after passing through three stationary points of energy and found to have three unstable modes, whose corresponding $\tilde\eta$-profiles are plotted in Fig. \ref{fig:3 modes}(b). From top to bottom, their growth rates are $\lambda \approx 0.034705, 0.026087$, and $0.013585$ respectively. In Fig. \ref{fig:linear compare2}, we shown the evolution of the perturbation $\tilde\eta$ which is initially set to be proportional to the most unstable mode (left) and the second most unstable mode (right). The black curves are the envelopes of $\tilde \eta$ calculated from linear stability analysis. It is clear that linear theory gives an excellent estimation for the most unstable mode but overestimates the growth rate of the second most unstable mode.

\subsubsection{Head-on collision}
\begin{figure}[!h]
    \centering
    \subfigure[]{
    \begin{minipage}{0.48\textwidth}
    \centering\includegraphics[width=\textwidth]{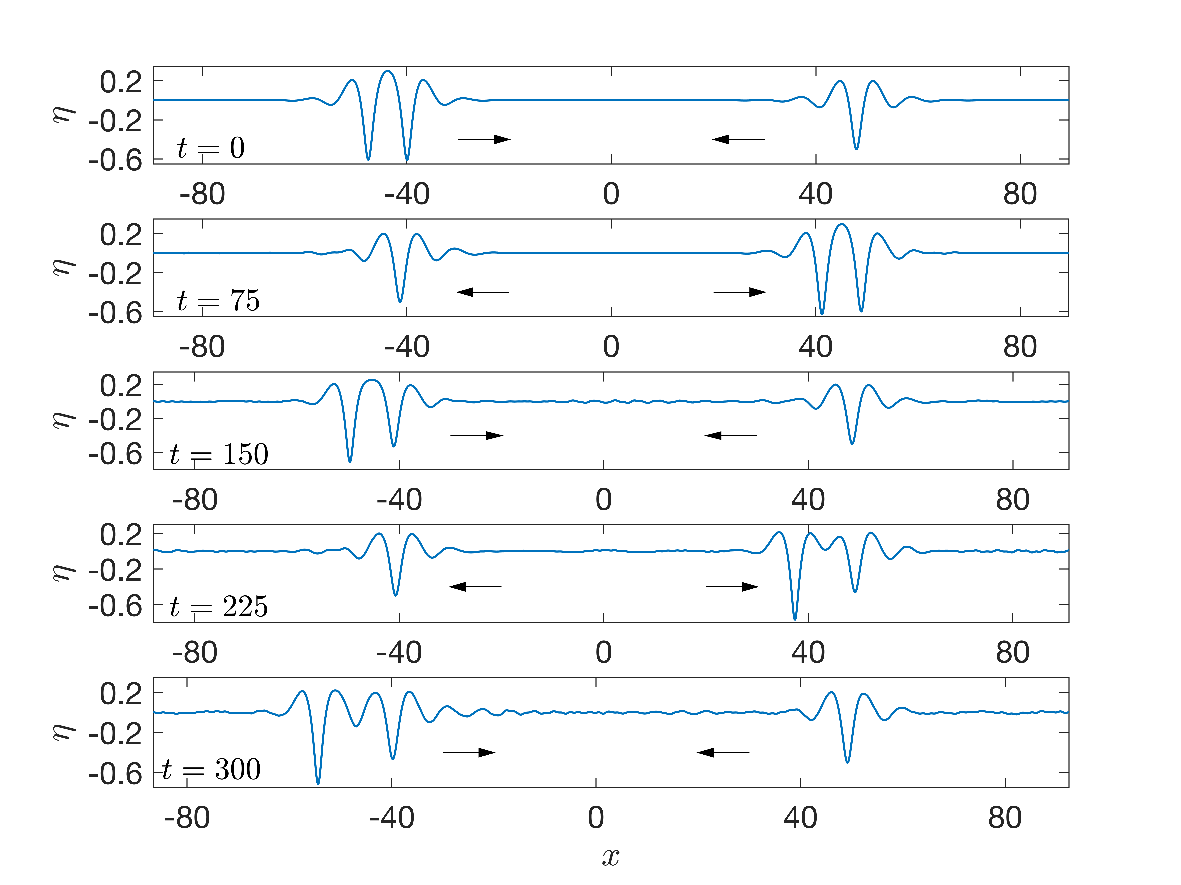}
    \end{minipage}
    }
    \subfigure[]{
    \begin{minipage}{0.48\textwidth}
    \centering\includegraphics[width=\textwidth]{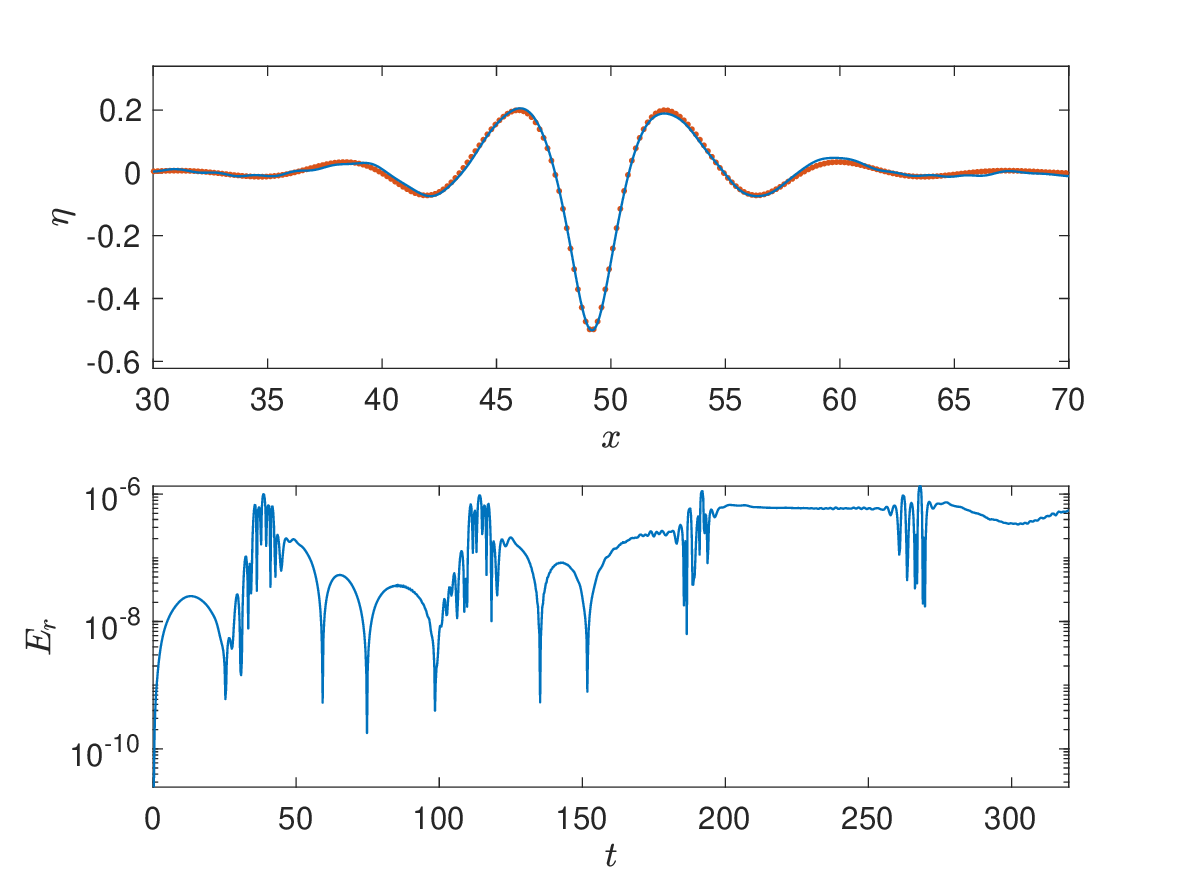}
    \end{minipage}
    }
    \caption{Head-on collisions between two solitary waves. (a) Profiles at $t = 0, 75, 150, 225$ and $300$. The arrows represent the direction of propagation of the solitary waves. (b) Top: comparison of the depression solitary wave at $t=0$ (red dots) and $t=300$ (blue curve). Bottom: $E_r$ during simulation.}
    \label{fig:collide1_R_0.2}
\end{figure}

We further check the stability of solitary waves by simulating their head-on collisions. In Fig. \ref{fig:collide1_R_0.2}, we choose two well-separated solitary waves as the initial condition of fully nonlinear simulation. One of them is of the depression type ($\eta(0) = -0.5$) and linearly stable, the other one is of the elevation type ($\eta(0) = 0.3$) and linearly unstable. We set $N=2048$ and $\Delta t = 5\times 10^{-3}$. Since we impose periodic boundary conditions, the two solitary waves can keep colliding during simulation. In Fig. \ref{fig:collide1_R_0.2}, we plot the initial profile ($t=0$) and other four typical profiles at $t = 75, 150, 225$, and $300$. The depression solitary wave is rather robust and maintains its shape after four collisions. This can be clearly seen from a comparison between its initial profile and the profile at $t=300$, which is shown on the top of Fig. \ref{fig:collide1_R_0.2}(a). The two profiles are almost indistinguishable except the tiny difference due to the radiated disturbances. On the other hand, the elevation solitary wave gradually loses its symmetry after two collisions with a manifestation that its two troughs have different heights. After the fourth collision, the elevation solitary wave undergoes complete distortion and begins to decompose into two side-by-side depression solitary waves. During the simulation, the energy error $E_r$ is  well controlled and less than $2\times 10^{-6}$. This is shown on the bottom of Fig. \ref{fig:collide1_R_0.2}(a).

\begin{figure}[!h]
    \centering
    \subfigure[]{
    \begin{minipage}{0.48\textwidth}
    \centering\includegraphics[width=\textwidth]{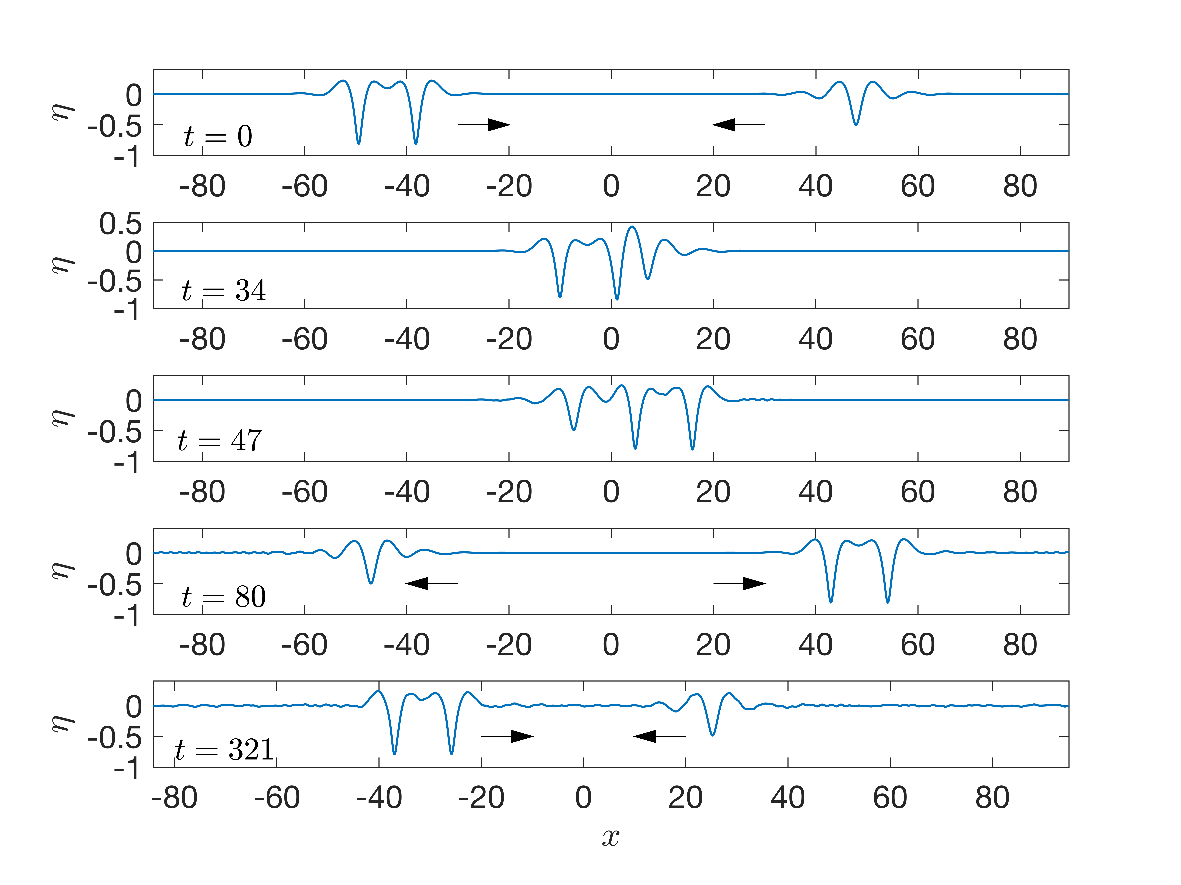}
    \end{minipage}
    }
    \subfigure[]{
    \begin{minipage}{0.48\textwidth}
    \centering\includegraphics[width=\textwidth]{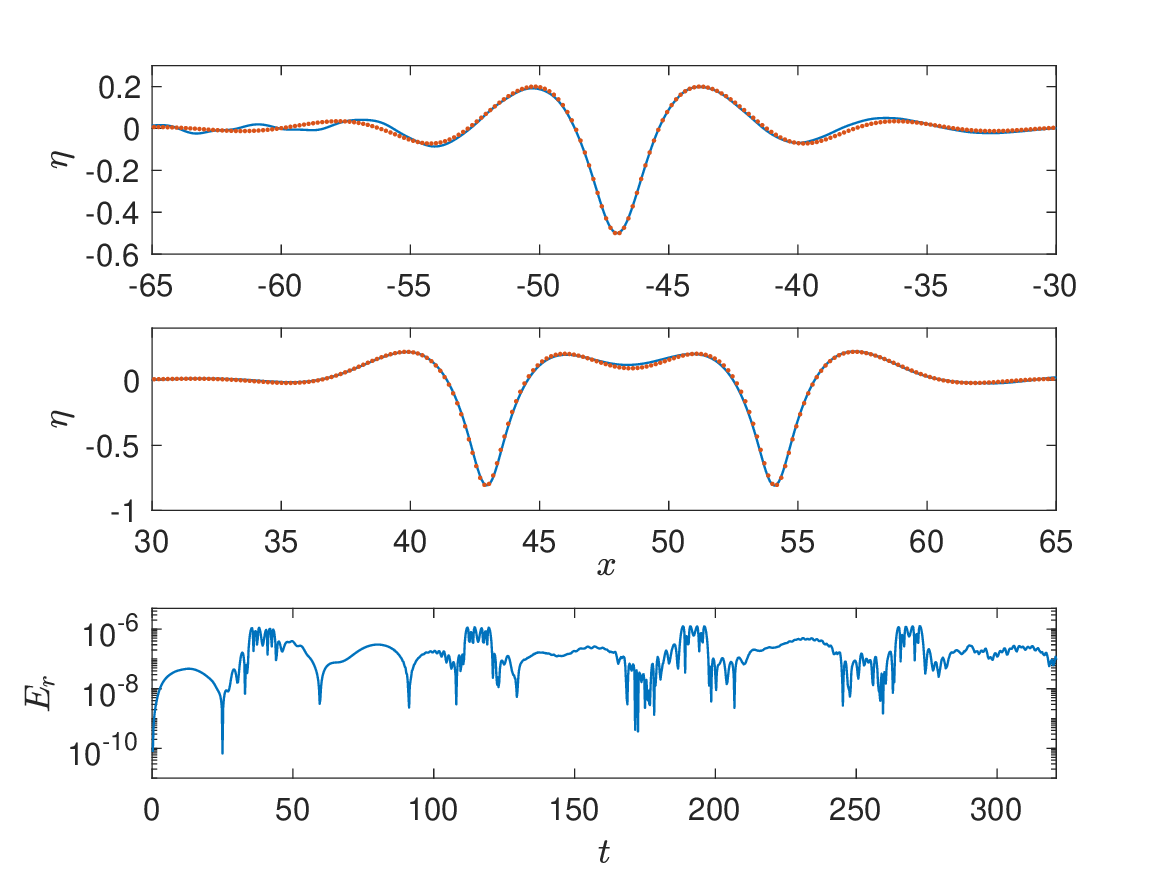}
    \end{minipage}
    }
    \caption{Head-on collision between two solitary waves. (a) Profiles at $t = 0, 34, 47, 80$ and $321$. The arrows represent direction of propagation of the solitary waves. (b) Top: comparison of the depression solitary wave at $t=0$ (red dots) and $t=321$ (blue curve). Middle: comparison of the elevation solitary wave at $t=0$ (red dots) and $t=321$ (blue curve). Bottom: $E_r$ during simulation.}
    \label{fig:collide2_R_0.2}
\end{figure}

In Fig. \ref{fig:collide2_R_0.2}, we display another simulation of head-on collisions between a depression solitary wave ($\eta(0) = -0.5$) and an elevation solitary wave ($\eta(0) = 0.09107$). Both waves are linearly stable according to linear stability analysis. We set $N=2048$ and $\Delta t = 5\times 10^{-3}$. The initial condition is shown on the top of Fig. \ref{fig:collide2_R_0.2}. The profiles at $t = 34, 47$, and $80$ show the process of the first collision and the separation later. Simulation continues until the fourth collision occurs. As expected, both solitary waves are rather robust and maintain their shape after collisions, as can be seen from Fig. \ref{fig:collide2_R_0.2} where the comparisons of their wave profiles are presented. There is only tiny difference between the initial profiles and the profiles at $t = 321$ for both solitary waves. On the bottom of Fig. \ref{fig:collide2_R_0.2} (b), we plot the energy error $E_r$ which is less the $2\times 10^{-6}$ during the simulation.

\subsection{Vortex roll-up}
Small-amplitude interfacial waves are neutrally stable according to the dispersion relation (\ref{c_p}). When there is a background shear across the interface, however, K-H instability will be triggered and generate the vortex roll-up structure \cite{hou1994removing}. Considering two uniform background currents $U_1$ and $U_2$ in the lower and upper layer, we have the following linear solutions
\begin{align}
    \eta(x,t) &= \text{Re}\big\{\epsilon \mathrm e^{\mathrm i(kx-\omega t)}\big\},\label{KH1}\\
    \phi_1(x,y,t) &= \text{Re}\big\{\mathrm i\epsilon \big(U_1-\omega/k\big)\mathrm e^{\mathrm i(kx-\omega t)}\mathrm e^{ky}\big\} + U_1x,\\
    \phi_2(x,y,t) &= \text{Re}\big\{\mathrm i\epsilon \big(\omega/k - U_2\Big)\mathrm e^{\mathrm i(kx-\omega t)}\mathrm e^{-ky}\big\} + U_2x,\label{KH3}
\end{align}
where $\omega$ has two complex roots given $R$ and $k$
\begin{align}
    \omega_{\pm} = \frac{k(U_1+RU_2)\pm\sqrt{|k|(1-R^2)+|k|^3(1+R)-Rk^2(U_2-U_1)^2}}{1+R}.
\end{align}
The critical condition to trigger the K-H instability is
\begin{align}
    |k|(1-R^2)+|k|^3(1+R)-Rk^2(U_2-U_1)^2<0.
\end{align}
Thus the shear must be strong enough to satisfy
\begin{align}
    (U_2-U_1)^2>2\sqrt{(1-R^2)(1+R)}/R.
\end{align}
Taking $\epsilon \ll 1$, we can replace the variable $x$ by $s$ in Eqs. (\ref{KH1})-(\ref{KH3}). Thus we can choose the following initial condition for our simulations
\begin{align}
    \eta(s,0) &= \epsilon \cos(ks),\\
    \varphi(s,0) &= \phi_1(s,0,0)-R\phi_2(s,0,0).
\end{align}
\begin{figure}[!h]
    \centering
    \subfigure[]{
    \begin{minipage}{0.48\textwidth}
    \centering\includegraphics[width=\textwidth]{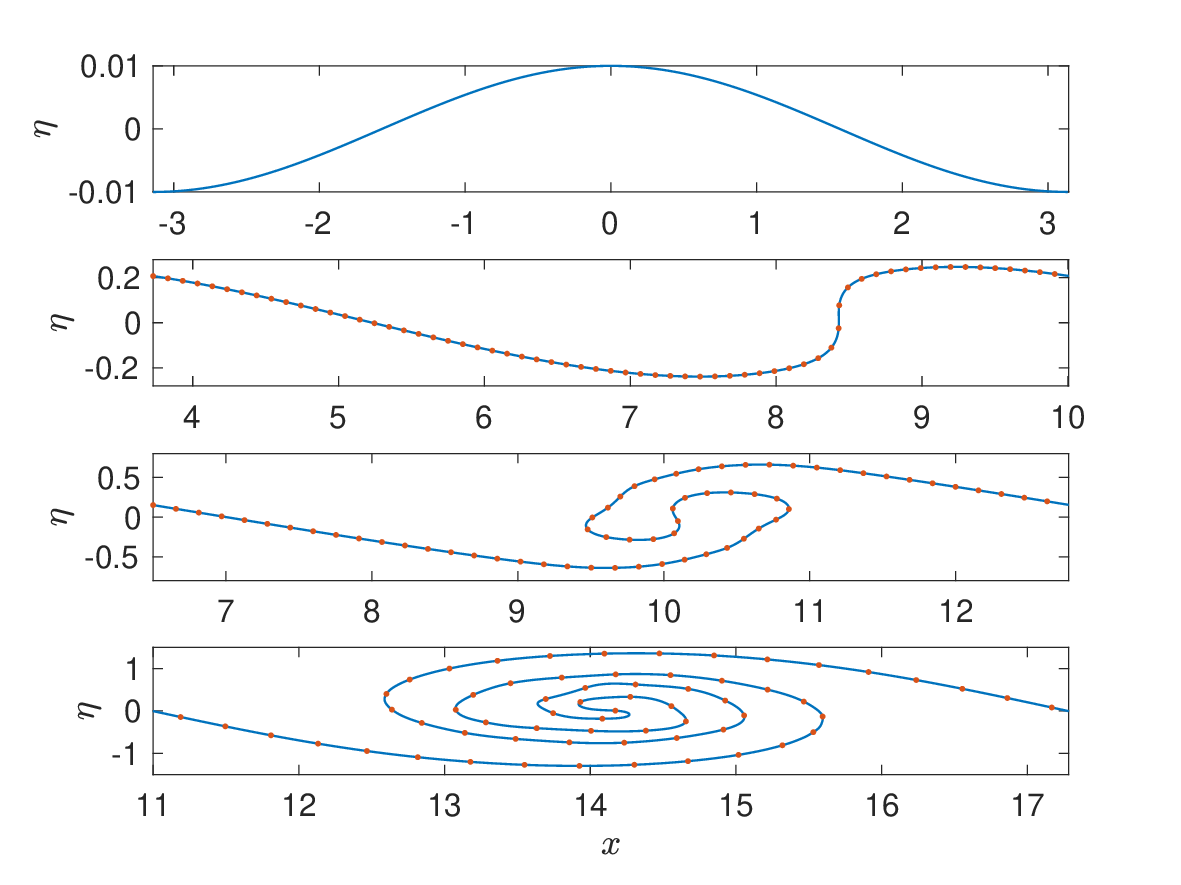}
    \end{minipage}
    }
    \subfigure[]{
    \begin{minipage}{0.48\textwidth}
    \centering\includegraphics[width=\textwidth]{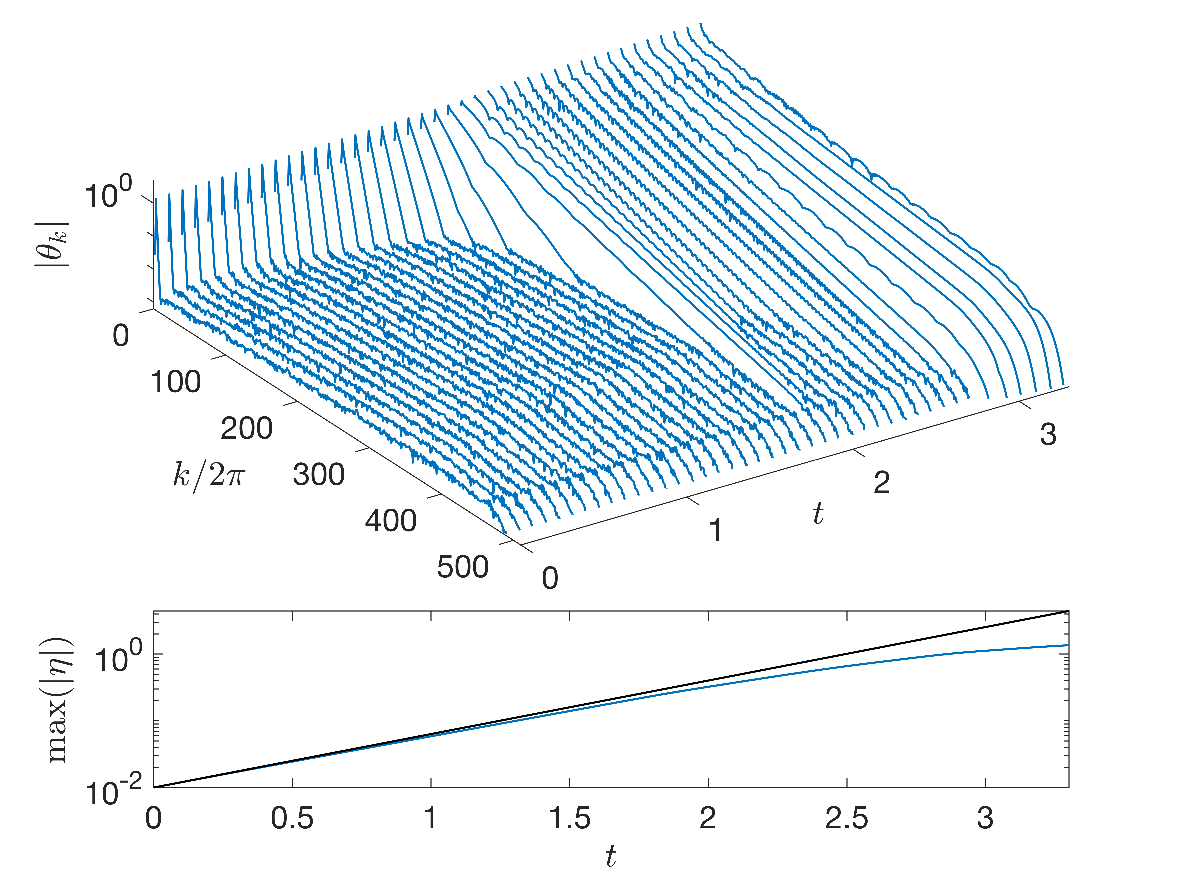}
    \end{minipage}
    }
    \caption{Simulation of K-H instability. (a) Profiles at $t = 0, 1.83, 2.5$, and $3.3$ from top to bottom. The red dots and blue curves represent solutions with $N = 512$ and $1024$. (b) Top: $|\theta_k|$ versus $k/2\pi$ at different moments. Bottom: max($|\eta|$) at different moments (blue), compared with linear theory (black).}
    \label{fig:KH}
\end{figure}
Fig. \ref{fig:KH}(a) displays a simulation of the vortex roll-up structure at different moments. We set $\epsilon = 0.01, R = 0.9$, $U_1 = 2, U_2 = -2, \Delta t = 10^{-4}$, and choose the unstable root $\omega_+ \approx 0.1053 + 1.8466\mathrm i$. The interface grows quickly in amplitude and follows the linear theory closely. At $t\approx 1.83$ it develops a vertical tangent. After that, the interface rolls over and forms two spiral fingers and grows in length. Close to the tip of these fingers, there is trace of small-amplitude wiggles due to the surface tension effect. This has also been previously observed and reported in \cite{hou1994removing,baker1998stable}. At $t=3.3$, the interface develops a standard vortex roll-up structure of the K-H type. As time progresses, the interface spirals, eventually leading to self-intersection, as reported in \cite{hou1994removing}. On the top of Fig. \ref{fig:KH}(b), we plot the spectrum of $\theta$ to show the effectiveness of the $36$th-order Fourier filter in suppressing the aliasing error. The bottom figure illustrates max$(|\eta|)$ against $t$ from both the fully nonlinear simulation (blue curve) and the linear theory (black curve). These curves closely align in a logarithmic scale until $t>2$.

\section{Conclusions}
In this paper, we present a novel boundary-integral method designed to simulate nonlinear two-dimensional unsteady interfacial waves and surface waves. The essential parts of the method are: (1) parameterizing the interface or surface by using their physical arclength, (2) establishing the Eulerian or mixed Eulerian-Lagrangian description, (3) using Cauchy's integral formula for solving the Laplace equation, and (4) employing the $\theta-s$ formulation to integrate in time and construct the interface. The advantages of our algorithm are due to the intrinsic dimension-reducing nature of the boundary-integral method, the numerical efficiency of the $\theta-s$ formulation, and the uniform distribution of sample points on the interface or surface. We are able to simulate various highly nonlinear wave motions, such as large-scale wave breaking, collisions of solitary waves, vortex roll-up structures due to the K-H instability, etc. We find that our algorithm possesses excellent numerical accuracy and robustness, making it very suitable for long-term simulations.

Particularly, we focus on the stability of symmetric interfacial gravity-capillary solitary waves in deep water. We calculate the depression and elevation solitary waves by using the Newton-Krylov method which can speed up the computation significantly. Previous studies on surface gravity-capillary solitary waves have shown that the depression waves are always stable. On the other hand, the elevation waves are unstable except those solutions between the first and the second stationary points on the energy curve. We reformulate the previous linear stability analysis and derive a new formulation suitable for interfacial waves. It possesses both numerical efficiency and robustness. This new formulation is employed to a special case with $R=0.2$ and yields that interfacial solitary waves have very similar stability properties to their counterpart of surface type. However, the energy curve of the depression solitary waves is not monotonic, resulting in exchanges of stability at the stationary points. Therefore, the depression solitary waves are linearly stable except a little part of them between the two stationary points of energy. For the elevation solitary waves, they are linearly unstable in small amplitude but turn to be stable after the first energy stationary point. We find that before the appearance of the second stationary point, elevation solitary waves are detected instability, which is a dominant superharmonic instability. At the second stationary point of energy, the exchange of stability happens again and solutions are linearly unstable thereafter. These findings are confirmed by our fully nonlinear simulations of solitary waves, especially their head-on collisions.

\section*{Acknowledgements}
X.G. would like to acknowledge the support from the Chinese Scholarship Council (no. 202004910418).

\appendix
\begin{appendices}

\section{Derivation of Eqs. (\ref{DdeltS}) and (\ref{DthetaDs})}\label{appendix:Ddeltas}
\begin{figure}[h!]
    \centering
    \subfigure[]{
    \begin{minipage}{0.45\textwidth}
    \centering\includegraphics[width=\textwidth]{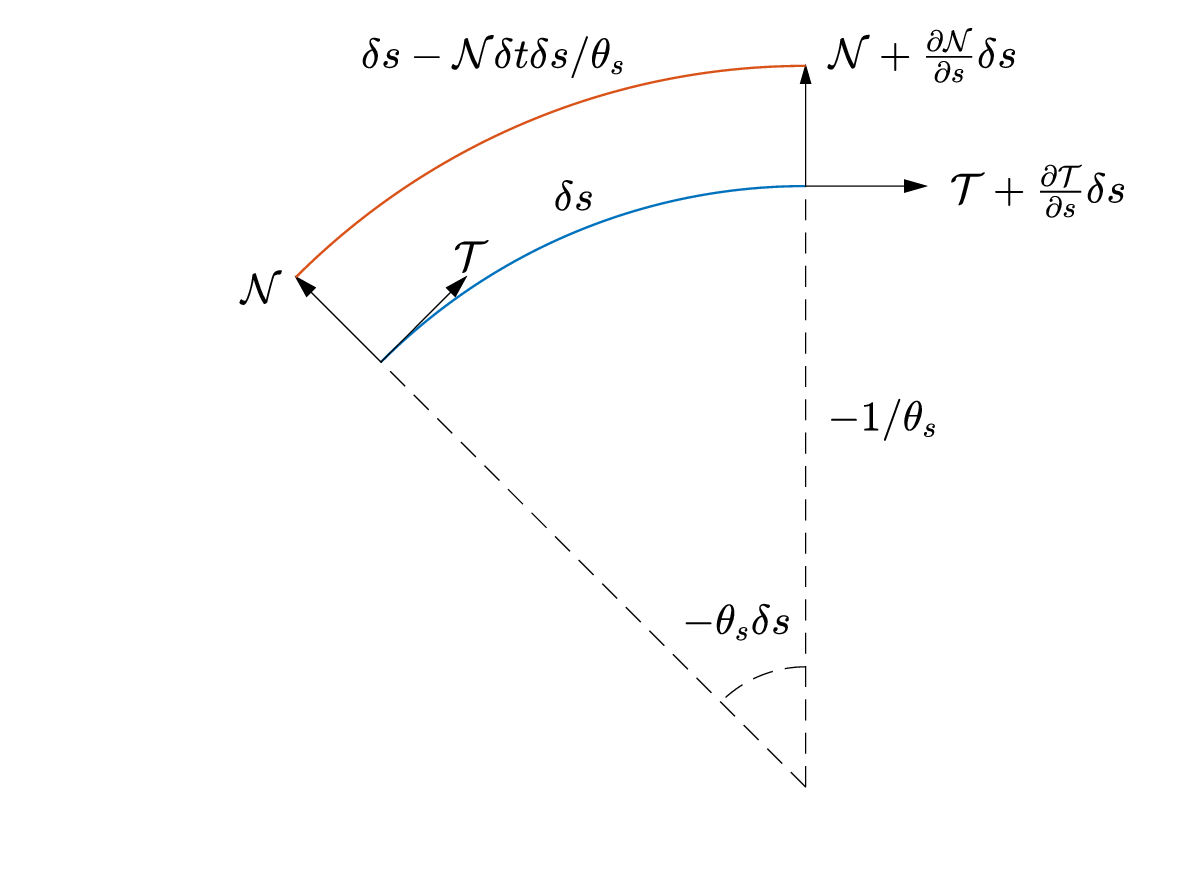}
    \end{minipage}
    }
    \subfigure[]{
    \begin{minipage}{0.45\textwidth}
    \centering\includegraphics[width=\textwidth]{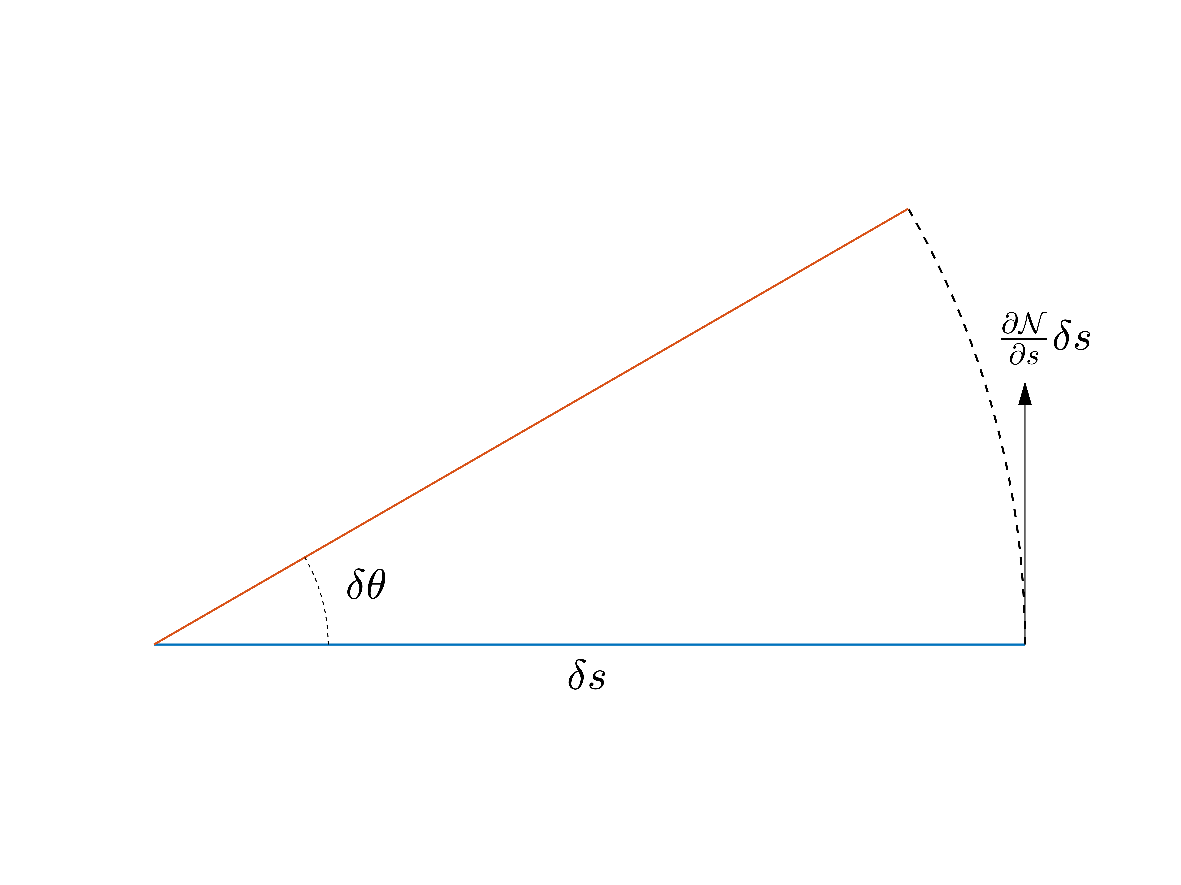}
    \end{minipage}
    }
    \caption{(a) A schematic of infinitesimal surface element (blue curve). The red curve illustrates the expansion effect due to the normal velocity $\mathcal N$ in $\delta t$ time. (b) An illustration of the rotational effect due to the normal velocity $\mathcal N$.}
    \label{fig:ds_dtheta}
\end{figure}
In Fig. \ref{fig:ds_dtheta}(a), we show an infinitesimal surface element with length $\delta s$ and curvature $\theta_s$. The increment of its length in $\delta t$ time consists of the contribution of the tangential velocity $\mathcal T$ and the normal velocity $\mathcal N$. The former stretches the element and generates an quantity $\partial \mathcal T/\partial s\delta s\delta t$. The normal velocity gives rise to an expansion effect, causing a quantity $-\mathcal N\delta t \delta s\theta_s$ in leading order. Note that the minus sign is due to the fact that $\theta_s<0$ is our schematic.
Putting these terms together, we have
\begin{align}
    \frac{D\delta s}{D t} = \big(\mathcal T_s - \theta_s\mathcal N\big)\delta s.
\end{align}
Similarly, the increment of inclination angle $\theta$ also comes from the contribution of $\mathcal T$ and $\mathcal N$. The former gives the element a translation in $\mathcal T$ direction, which causes a quantity $\mathcal T\delta t \theta_s$. The normal velocity generates a local rotation with angle $\partial \mathcal N/\partial s \delta t$ (see Fig. \ref{fig:ds_dtheta} (b)). Thus we have
\begin{align}
    \frac{D\theta}{Dt} = \mathcal N_s + \theta_s \mathcal T.
\end{align}

\section{Derivation of $\gamma(t)$ in Case II}\label{appendix:gamma}
\begin{figure}[h!]
    \centering
    \includegraphics[width=0.65\textwidth]{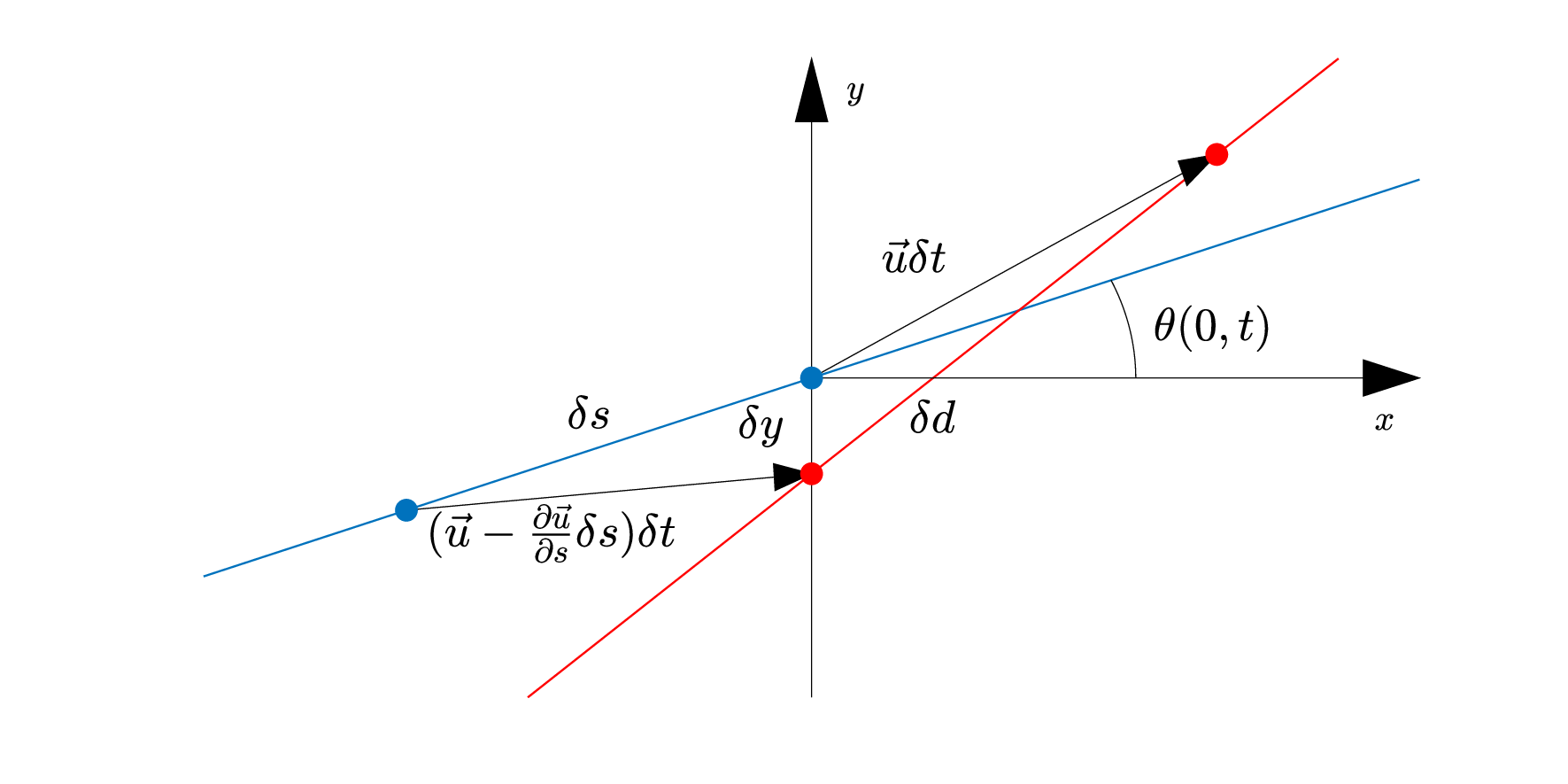}
    \caption{A local schematic of interface and two sample particles at $t$ (blue) and $t+\delta t$ (red).}
    \label{fig:gamma}
\end{figure}

Fig. \ref{fig:gamma} shows a local schematic of an interface and two Lagrangian particles on it at time $t$ (blue) and $t+\delta t$ (red). Initially, the distance between the two particles is $\delta s$ and the right one is located on the left boundary of the computational domain, which is set to be the the $y$-axis without loss of generality. In $\delta t$ time, the left particle moves to the boundary. According (\ref{DsDt})
\begin{align}
    \gamma(t) = \frac{D s}{Dt}\Big|_{s = 0} = \lim_{\delta t\rightarrow 0} \frac{\delta d}{\delta t},\label{gamma(t)}
\end{align}
where $\delta d$ is the distance between the two red points. Using the geometric relation, we have
\begin{align}
    \delta d = \sqrt{(u(0,t)\delta t)^2 + (v(0,t)\delta t - \delta y)^2}.\label{delta d}
\end{align}
The minus sign of $\delta y$ is due to the fact that we define $\delta y$ as the vertical component of the vector starting from the blue point and pointing to the red point on the $y$-axis, which is negative in the schematic. 
On the other hand, we have
\begin{align}
    \delta s \cos \theta(0,t) &= \Big(u(0,t) + \frac{\partial u}{\partial s}(0,t)\delta s\Big)\delta t,\\
    \delta s \sin \theta(0,t) &= \Big(v(0,t) + \frac{\partial v}{\partial s}(0,t)\delta s\Big)\delta t - \delta y.
\end{align}
In the leading order, we obtain
\begin{align}
    \delta y = \Big(v(0,t) - u(0,t)\tan\theta(0,t)\Big)\delta t. \label{delta y}
\end{align}
Note that this is equivalent to the kinematic boundary condition at $x=0$. Substituting (\ref{delta y}) into (\ref{delta d}) and (\ref{gamma(t)}), we have
\begin{align}
    \gamma(t) = \frac{u(0,t)}{\cos\theta(0,t)},
\end{align}
or equivalently 
\begin{align}
    \gamma(t) = \mathcal T(0,t) - \mathcal N(0,t)\tan\theta(0,t).
\end{align}

\end{appendices}

\bibliographystyle{plain}
\bibliography{references}

\end{document}